\documentclass[preprint,12pt,number]{elsarticle}

\usepackage{amssymb}

\usepackage{amsmath}
\usepackage{graphicx}
\usepackage{array}
\usepackage{ragged2e}
\usepackage{booktabs}
\usepackage{geometry}
\usepackage{longtable}
\usepackage{multirow}
\usepackage{hyperref}
\usepackage[numbers]{natbib}
\journal{}

\begin{document}

\begin{frontmatter}



\title{Large Language Models integration in Smart Grids}


\author[inst1]{Seyyedreza Madani}

\author[inst1]{Ahmadreza Tavasoli}

\author[]{Zahra Khoshtarash Astaneh}

\author[inst1,inst3]{Pierre-Olivier Pineau}

\affiliation[inst1]{organization={Chair in Energy Sector Management, HEC Montreal},
            addressline={3000 Chem. de la Côte-Sainte-Catherine}, 
            city={Montreal},
            postcode={H3T 2A7}, 
            state={Quebec},
            country={Canada}}


\affiliation[inst3]{organization={Decision Sciences, HEC Montreal},
            addressline={3000 Chem. de la Côte-Sainte-Catherine}, 
            city={Montreal},
            postcode={H3T 2A7}, 
            state={Quebec},
            country={Canada}}


\begin{abstract}
Large Language Models (LLMs) are changing the way we operate our society and will undoubtedly impact power systems as well—but how exactly? By integrating various data streams—including real-time grid data, market dynamics, and consumer behaviors—LLMs have the potential to make power system operations more adaptive, enhance proactive security measures, and deliver personalized energy services. This paper provides a comprehensive analysis of 30 real-world applications across eight key categories: Grid Operations and Management, Energy Markets and Trading, Personalized Energy Management and Customer Engagement, Grid Planning and Education, Grid Security and Compliance, Advanced Data Analysis and Knowledge Discovery, Emerging Applications and Societal Impact, and LLM-Enhanced Reinforcement Learning. Critical technical hurdles, such as data privacy and model reliability, are examined, along with possible solutions. Ultimately, this review illustrates how LLMs can significantly contribute to building more resilient, efficient, and sustainable energy infrastructures, underscoring the necessity of their responsible and equitable deployment.
\end{abstract}

\begin{graphicalabstract}
\includegraphics[width=\textwidth]{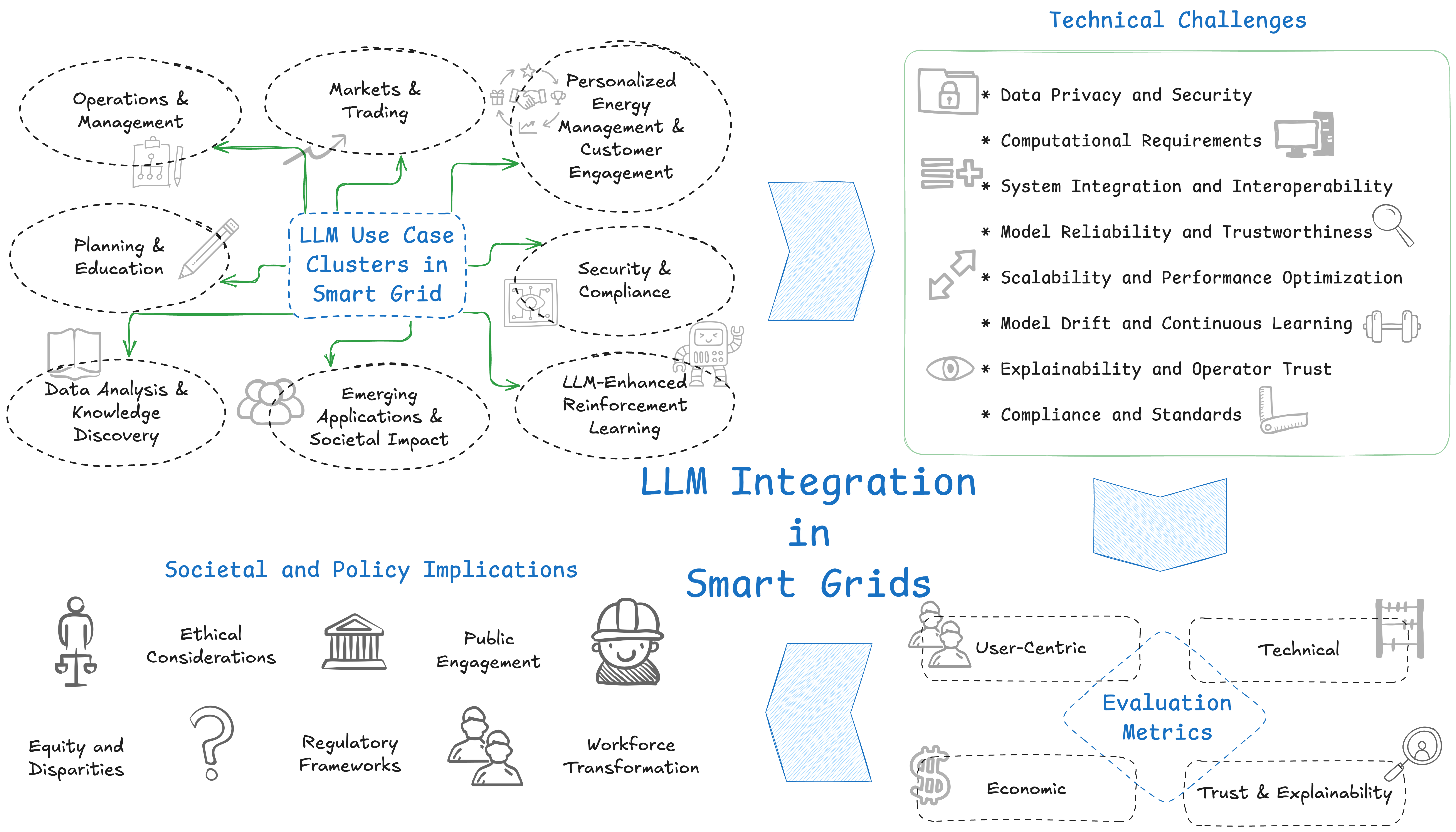}
\end{graphicalabstract}

\begin{highlights}
\item LLMs can enhance smart grid operations through  natural language capabilities.

\item AI-driven analysis can improve grid security and reliability.

\item LLMs could personalize energy management and empower consumers.

\item Ethical AI integration is crucial for equity and trust in smart grids.

\item LLMs can unlock new optimization opportunities via reinforcement learning.
\end{highlights}


\end{frontmatter}


\section{Introduction}
\label{sec:intro}
The global energy sector is rapidly changing, driven by the rapid development of small-scale generation and storage technologies and the need for sustainable, efficient, and resilient power. This shift is transforming traditional power grids into smart grids – advanced, digital networks \citep{bayindir2016smart}. Smart grids differ from their predecessors by being distributed, having two-way communication, and integrating various energy sources. This evolution is vital for incorporating more renewable energy like solar and wind, meeting growing demands for reliable energy, and improving security in our interconnected world. Smart grids have the potential to offer greater flexibility, quicker responses, and better sustainability, but they also present new obstacles.

A major challenge is the inherent complexity of smart grids \citep{koliba2014salience}. Managing numerous interconnected devices like smart meters and electric vehicle chargers is difficult. Adding to this are fluctuating energy demands due to weather and societal factors \citep{staffell2018increasing, torriti2017understanding}. Furthermore, integrating intermittent renewable sources like solar and wind makes maintaining grid stability and balancing supply with demand harder \citep{koliba2014salience}. Traditional grid management struggles with this dynamic complexity.

Scalability is another key concern. As electricity dependence increases, grids must expand to meet demand reliably and efficiently. This requires not just more infrastructure, but also intelligent control systems to manage a growing number of devices and resources \citep{rodriguez2018scalability}. Decentralization through microgrids and prosumers (those who both consume and generate energy) further complicates scalability. Existing centralized management systems are not designed for these diverse scenarios, requiring more adaptable technologies.

Smart grids also face new security threats due to their reliance on digital technologies \citep{bekara2014security}. Cyberattacks could disrupt operations, causing outages, economic losses, and safety risks. Robust security measures, constant monitoring, and effective incident response are crucial to protect the grid. Enhancing both physical and cyber security is a priority for modern energy systems.

To address these challenges, Artificial Intelligence (AI), particularly Large Language Models (LLMs), offers powerful solutions.  LLMs are advanced AI systems trained on extensive corpora of text, code, image, sound and video, enabling them to process and generate natural language with remarkable depth and accuracy. This capacity allows LLMs to analyze complex data, extract insights from unstructured information, and communicate with human operators more intuitively. Recent research has successfully demonstrated LLMs' capability to extract critical asset information efficiently from diverse sources, significantly enhancing decision-making in complex operational environments like the oil and gas industry \citep{chen_advancing_2025}. Moreover, LLMs can interpret context from various sources like sensor data and weather reports, identifying important relationships that might otherwise go unnoticed.

Moreover, LLMs can aid in decision-making by generating actionable insights and clear recommendations. This builds trust among operators, enabling better decisions in dynamic situations.The integration of local knowledge into LLMs further enriches their capabilities, significantly enhancing societal comprehension and engagement in achieving carbon neutrality \citep{han_integrating_2024}. LLMs act as intelligent intermediaries, translating human instructions into actions and providing clear justifications.

This paper contributes to a growing body of literature reviewing emerging technologies in the energy sector, such as blockchain \citep{andoni2019blockchain}, deep learning \citep{massaoudi2021deep}, and broader artificial intelligence techniques \citep{omitaomu2021artificial}. Opportunities also exist for integrating LLMs in engineering design processes, automating creativity and reasoning tasks traditionally performed by humans \citep{gopfert_opportunities_2024}. Similarly, we identify key opportunities, technical challenges, and implementation scenarios specifically for LLMs.

We investigate diverse possible use cases of LLMs in smart grids, highlighting their potential to enhance grid operations, optimize energy markets, empower consumers, and bolster cybersecurity. By organizing these use cases into eight distinct clusters, we provide an overview of how LLMs can effectively address critical challenges within power systems, delivering targeted benefits ranging from operational efficiency improvements to enhanced security.

Our analysis demonstrates that LLMs could significantly contribute to creating more efficient, reliable, and sustainable energy infrastructures. Additionally, we illustrate current capabilities and explore potential future roles of LLMs, emphasizing the importance of ongoing research and innovation in meeting emerging challenges within the energy sector. Ultimately, this work aims to inform and inspire further exploration at the intersection of artificial intelligence and energy, fostering a resilient, equitable, and environmentally responsible energy future.

Following this introduction, Section \ref{sec:literature} reviews relevant literature. Section \ref{sec:clusters} details use cases, organized into eight clusters. Section \ref{sec:challenges} examines the technical challenges of deploying LLMs and potential solutions. Section \ref{sec:evals} outlines key evaluation metrics. Section \ref{sec:soc} discusses societal and policy implications. Section \ref{sec:future} explores future directions, and Section \ref{sec:conclusion} concludes the paper.

\section{Literature Review}
\label{sec:literature}

This literature review begins by briefly exploring smart grids, highlighting how digital transformation has enhanced operational efficiency while also introducing new security concerns. It then positions LLMs within this evolving energy landscape, emphasizing how their sophisticated natural language processing and data-driven reasoning capabilities can address vulnerabilities and streamline grid management.

Smart grid modernization, driven by digital and information technologies, features advanced metering and communication \citep{Omitaomu2021}. This progress, while beneficial, creates data management and security challenges. Early work by \cite{Casagrande2014} explored using Natural Language Processing (NLP) for defining smart metering system goals, establishing the value of textual data in power system design.

The increasing sophistication of cyber threats in the Industrial Internet of Things (IIoT), crucial for smart grids, raises concerns. \citet{Dong2021}'s research on text attacks against NLP models in smart grids highlights vulnerabilities. This emphasizes the need for robust security and awareness of NLP's susceptibility to attacks. \citet{Jha2023} further investigated the intersection of machine learning (ML) and NLP for enhanced smart grid cybersecurity, reviewing techniques for risk analysis and anomaly detection to improve defenses. The advancements in NLP, especially with machine learning, are largely due to the rise of LLMs.

LLMs represent a significant leap in AI, driven by neural network evolution and large datasets. The Transformer architecture \citep{Vaswani2017}, with its attention mechanisms, enabled more effective processing of sequential data compared to previous recurrent networks. This allowed models to capture long-range text dependencies, improving language understanding and generation. The extensive training of these models, involving trillions of parameters, significantly boosted their capabilities. Early models like BERT \citep{Devlin2018} demonstrated the effectiveness of pre-training on large text datasets for various tasks. This pre-training followed by fine-tuning became standard, allowing LLMs to adapt to diverse applications. The combination of the Transformer architecture, large-scale pre-training, and model scaling led to the emergence of LLMs as powerful tools.

Recently, LLMs have offered new solutions for smart grid challenges. Their ability to learn patterns in text and code enables them to perform tasks like text generation, summarization, translation, question answering, and code completion \citep{Brown2020}. Their understanding of context has led to adoption in areas from customer service \citep{Radford2019} and content creation to science and engineering \citep{Thirunavukarasu2023, Bubeck2023}. LLMs' ease of use has made tools like ChatGPT widely popular for information, content generation, workflow automation, and human-computer interaction. Their increasing accessibility and performance indicate growing impact. Several papers highlight their potential: \citet{Huang2023} explored GPT-4's capabilities in power system tasks like optimal power flow and EV scheduling. \citet{Hamann2024Foundation} advocate for grid-specific foundation models to leverage diverse grid data. \citet{Mohammadabadi2024} introduced generative AI for distributed learning in smart grids to improve communication and privacy.

LLMs are also enhancing human-machine interaction in power systems. \citet{Jin2024} presented ChatGrid, using ChatGPT for visualizing transmission networks through natural language queries. \citet{Na2024} explored Transformer models for improved speech recognition in grid dispatch. \citet{Maddigan2023} introduced Chat2VIS, using LLMs to generate data visualizations from natural language. \citet{Pribylskii2024} examined LLMs for aiding decision-making in distributed generation smart grids.

The security implications of LLMs in smart grids are under review. \citet{Ruan2024} highlighted potential threats and the need for countermeasures. \citet{Li2024} assessed LLM risks in smart grids, identifying data injection and knowledge retrieval attacks. \citet{Selim2024} validated LLMs for cyberattack detection on smart inverters using control command analysis. \citet{Zaboli2024Novel} explored LLMs for anomaly detection in smart grid communication, proposing a dialogue system for faster implementation. \citet{Zhang2024} reviewed attacks and defenses for machine learning in IoT smart grids. \citet{GATTAL2024} investigated LLMs for improving the interpretability of anomaly detection using SHAP values.

Beyond security, LLMs are used to process complex text in power grid operations. \citet{Lin2023} proposed a knowledge graph embedding model using pre-trained language models for power grid defect knowledge. \citet{Chen2023} presented a knowledge graph reasoning model using pre-training for decision-making in distribution network defects. \citet{Yang2024} focused on enhancing text generation for power grid maintenance strategies using fine-tuning. The role of semantic web technologies in smart grid information integration was studied by \citet{Zhou2012}, and tools for accessible information models were presented by \citet{Crapo2009}.

Ethical considerations of AI in smart grids are increasingly important. \citet{Hayawi2024} proposed ethical principles for smart grids, emphasizing transparency, accountability, fairness, and privacy.

Specific LLM applications in power system analysis are emerging. \citet{Zhao2024} explored LLMs for determining partial tripping of distributed energy resources. \citet{Liu2024} introduced Lfllm for short-term load forecasting. \citet{Jing2024} proposed combining LLMs with prompt engineering for fault diagnosis. \citet{Wang2024} investigated LLMs for predicting electrical safety incidents using smart meter data. The use of data analytics for renewable energy systems in Smart Grid 2.0 is reviewed by \citet{Nguyen2024}, who also addressed security. \citet{Shi2024} reviewed LLM adoption for smart grid application acceleration. Safe reinforcement learning in smart grids with high Distributed Energy Resource (DER) penetration is reviewed by \citet{Bui2024}. \citet{Wilker2017} compared toolboxes for standardized smart grid modeling, and \citet{Zibaeirad2024} surveyed smart grid security, including LLMs.

This growing literature shows significant interest in using LLMs for smart grid challenges. However, research gaps exist, including a lack of systematic application categorization, comprehensive analysis of technical challenges and solutions, robust evaluation frameworks, and exploration of societal and policy implications. This paper aims to address these gaps by categorizing LLM use cases, analyzing technical barriers and solutions, discussing evaluation metrics, and examining the societal and policy aspects of LLM integration in next-generation power systems.

\section{Clusters and Use Cases}
\label{sec:clusters}
Building upon the established importance of LLMs for smart grids, this section presents specific applications. To facilitate a clear and organized understanding of their potential, we detail 30 representative use cases identified and analyzed using a systematic approach. This involved a review of academic literature, industry reports, and practical implementations, followed by an evaluation based on relevance to smart grid needs, feasibility given current LLM capabilities, and potential for impactful innovation. This process ensured a balanced representation of how LLMs can contribute to intelligent energy infrastructures.

The subsequent presentation of these use cases is organized into eight distinct clusters, each representing a significant area of LLM application within the smart grid ecosystem. This categorization framework highlights the common functionalities, shared technological underpinnings, and user-centric focus within each group. By structuring the applications in this manner, we aim to provide a thematic understanding of how LLMs can be leveraged to address specific challenges and unlock targeted benefits, ranging from enhanced grid operations and optimized energy markets to consumer empowerment and improved grid security. This structured approach facilitates a deeper appreciation for the versatility and broad impact of LLMs in the future of energy.

\subsection{Cluster 1: Smart Grid Operations \& Management}

This cluster focuses on how LLMs can revolutionize the core operational aspects of smart grids. The use cases discussed in this section showcase LLMs as intelligent assistants, capable of enhancing grid management, anomaly detection, resilience assessment, fault localization, and microgrid control. By leveraging their natural language processing and reasoning capabilities, LLMs can offer more intuitive, dynamic, and efficient solutions compared to traditional methods, ultimately leading to improved grid reliability, stability, and performance.

\subsubsection{Natural Language-Driven Grid Topology Reconfiguration \& Optimization}

The complexity of modern smart grids, characterized by numerous interconnected devices and dynamic loads, makes manual reconfiguration a significant challenge. Operators require more intuitive methods to interact with the system, and LLMs are well-suited to bridge this gap. By understanding high-level, natural language instructions, LLMs act as an intelligent intermediary, translating human intentions into precise control commands. For example, an operator might instruct the LLM to "Prioritize power flow to the hospital district during the storm," or to "Balance load across substations on the west side." 

Traditional grid management relies on complex graphical user interfaces (GUIs) with numerous menus and parameters, often requiring extensive training and leading to potential human errors. Automated systems are usually rule-based and lack the flexibility to adapt to unforeseen situations. LLMs offer a more agile and intuitive interface, enabling faster response times, reducing training needs, and allowing operators to leverage their expertise more effectively. Additionally, an LLM can provide a natural language explanation of `why' it chose a particular reconfiguration, boosting operator confidence and understanding.  Implementation requires a robust real-time data pipeline that feeds grid topology data, performance metrics, and contextual information to the LLM. The LLM must be trained on a vast dataset of grid operations manuals, expert knowledge, and potentially simulation outputs. Furthermore, secure and reliable communication channels between the LLM and the grid control system are essential. Ensuring the LLM's decisions are safe and reliable is also a key challenge and necessitates rigorous testing before live deployment.

\subsubsection{Autonomous Grid Anomaly Detection \& Contextual Explanation}

Modern smart grids generate enormous data streams that make identification of genuine anomalies difficult. These anomalies signal potential problems such as equipment failures, cyberattacks, and unusual load patterns. LLMs can analyze these diverse data streams holistically, detecting subtle patterns and correlations that might be missed by traditional threshold-based or statistical methods. The LLM can then generate human-readable explanations of `why' an event is considered anomalous, linking it to relevant contextual information. 

Traditional anomaly detection systems often trigger numerous false positives and struggle with novel anomalies. LLMs, by understanding context and learning complex relationships, can reduce false positives and identify more subtle anomalies. Crucially, contextual explanations provide operators with the ability to assess the severity and potential impact of an anomaly, allowing them to prioritize responses efficiently. Implementation involves integrating LLMs with various data sources including SCADA systems (Supervisory Control and Data Acquisition systems that monitor and control industrial processes and equipment in real time), substation IED (Intelligent Electronic Devices) logs, cybersecurity monitoring tools, and external data like weather reports. Training requires feeding the LLM historical grid data, labeled anomalies with expert explanations, and system documentation. Developing methods to correlate information from different sources and generate coherent explanations is crucial, along with ensuring that these explanations are accurate and trustworthy.

\subsubsection{Predictive Grid Resilience Assessment with Scenario Narrative Generation}

Assessing the resilience of a smart grid against disruptions, such as natural disasters or cyberattacks, requires understanding complex interdependencies and cascading effects. LLMs can analyze historical grid performance data, infrastructure data, weather patterns, geographical information, and even social sentiment to predict the grid's vulnerability to specific disruptive events. The LLM can also generate narrative descriptions of potential failure cascades, making these complex scenarios understandable for planners and decision-makers.

Traditional resilience assessments often rely on probabilistic risk models and static analyses which fail to capture the dynamic nature of grid failures. LLMs can incorporate a wider range of factors and generate detailed and realistic scenarios, enabling stakeholders to visualize potential impacts. The narrative generation aspect is vital for communicating these risks to non-technical audiences and justifying investments in resilience measures. Implementation requires integrating LLMs with diverse data sources, including historical outage data, weather APIs, GIS data, asset management systems, and social media analysis tools. Training LLMs require data on past outages, expert analyses of cascading failures, and simulations of disruptive events. A significant challenge is the development of methods for the LLM to generate plausible and informative narratives that accurately reflect potential failure modes and their consequences.

\subsubsection{AI-Augmented Fault Localization \& Repair Workflow Orchestration}

When faults occur in the grid, quick identification of their location and the optimal repair procedure are critical for minimizing downtime. LLMs can analyze multiple information sources, including text-based fault reports, images of damaged equipment, sensor data, maintenance logs, and equipment manuals. Based on this analysis, the LLM suggests the most likely fault locations, proposes step-by-step repair procedures, and helps orchestrate the repair workflow by identifying necessary tools, parts, and personnel.

Traditional fault localization often relies on predefined procedures and technician expertise, which can be slow and prone to error, particularly for complex or unusual faults. LLMs can significantly speed up diagnostics by leveraging unstructured data and automating initial troubleshooting steps. By suggesting optimal repair workflows and resource allocation, LLMs improve efficiency, reduce repair times, and minimize outage impacts. Implementation requires seamless integration with asset management systems, maintenance reporting tools, sensor data platforms, and technician communication platforms. Training LLMs involve a large dataset of past fault reports, repair logs, equipment manuals, and potentially expert knowledge. A key challenge is enabling the LLM to understand visual information alongside textual and numerical data.

\subsubsection{Dynamic Microgrid Management with Natural Language Coordination}

Microgrids, with their diverse mix of DERs, require sophisticated control strategies to optimize energy flow, maintain stability, and ensure resilience. LLMs can act as an intelligent coordinator, interpreting high-level objectives from microgrid operators in natural language and translating these objectives into control commands for individual DERs. Operators can instruct the LLM to "Prioritize renewable energy usage while maintaining voltage stability," or to "Prepare for islanded operation due to incoming storm."

Traditional microgrid management systems often rely on pre-programmed rules or centralized control algorithms, which lack flexibility and adaptability. LLMs offer a more dynamic and context-aware approach, allowing for sophisticated optimization and faster adaptation to changing conditions. The natural language interface simplifies interactions for operators, making microgrid management more intuitive.  Implementation requires a robust communication infrastructure to connect the LLM with individual DER control systems. The LLM must be trained on the operational characteristics of DERs, microgrid control principles, communication protocols, and optimization objectives. Developing secure communication channels and ensuring that the LLM's commands are executed safely is also crucial.

\subsection{Cluster 2: Smart Energy Markets \& Trading}

This cluster highlights the transformative potential of LLMs in smart energy markets and trading. The use cases presented demonstrate how LLMs can automate complex processes, enhance market intelligence, facilitate decentralized trading, and optimize pricing strategies. By leveraging their abilities in natural language processing, data analysis, and decision-making, LLMs can foster more efficient, transparent, and dynamic energy markets, leading to better outcomes for both producers and consumers.

\subsubsection{Autonomous Power Purchase Agreement Negotiation \& Drafting with Contextual Awareness}

The negotiation and drafting of Power Purchase Agreements (PPAs) is often complex, time-consuming, and requires significant legal and financial expertise. LLMs can automate significant portions of this process by analyzing market conditions, regulatory requirements, the specific needs of the buyer and seller, and even historical PPA terms. Based on a comprehensive analysis, an LLM can autonomously negotiate terms and draft customized PPAs that are fair and mutually beneficial.

Traditional PPA negotiation involves lengthy discussions between legal teams, often missing optimal terms or taking significant time. LLMs can significantly streamline this process by identifying optimal pricing, contract duration, and other terms more efficiently than human negotiators. They can also provide transparent explanations of the rationale behind the suggested terms, fostering trust and understanding. The implementation requires training an LLM on a vast corpus of historical PPAs, regulatory documents, market data, and expert knowledge. Integration with contract management systems and secure negotiation platforms is essential. A crucial challenge lies in ensuring the LLM adheres to legal and ethical guidelines and that the generated agreements are legally sound.

\subsubsection{Energy Market Sentiment Analysis \& Predictive Trading Strategies}

Energy markets are influenced by a complex interplay of factors, including supply and demand fundamentals, geopolitical events, technological advancements, and public sentiment. LLMs can process vast amounts of unstructured data like news articles, social media posts, financial reports, and analyst commentaries to gauge market sentiment towards various energy commodities. By identifying patterns and trends in this sentiment, LLMs can generate insights and inform more sophisticated trading strategies.

Traditional quantitative trading models primarily rely on historical price data and economic indicators. LLMs can capture more nuanced and real-time information from unstructured sources, predicting short-term price fluctuations and identifying emerging market trends that traditional models might miss. The ability to understand the sentiment behind market commentary can be a valuable predictive signal. Implementation requires access to real-time news feeds, social media APIs, financial data sources, and analyst reports. Training the LLM involves natural language processing techniques for sentiment analysis, topic modeling, and named entity recognition. Integrating the LLM’s insights with algorithmic trading platforms requires careful consideration of risk management and regulatory compliance.

\subsubsection{Decentralized Energy Trading Platform with Natural Language Offer Matching}

Existing energy markets are often centralized and dominated by large players. A decentralized energy trading platform, leveraging LLMs, could empower smaller-scale energy producers and consumers to directly trade energy with each other. LLMs can facilitate this by understanding natural language offers and requests. For example, someone might offer to sell 5 kWh of solar energy between 2 PM and 4 PM at \$0.10/kWh, while someone else might want to buy 10 kWh of renewable energy that evening. The LLM would act as an intelligent matchmaker, connecting compatible buyers and sellers.

This approach bypasses intermediaries, potentially leading to better prices for both producers and consumers, fostering a more localized and resilient energy system. The natural language interface makes it easier for individuals without specialized knowledge to participate in the market. Implementation requires a secure and transparent blockchain-based platform for recording energy transactions. The LLM must be trained on the semantics of energy trading terms, pricing structures, and different types of energy sources. Integration with smart meters for accurate measurement is essential. Addressing regulatory hurdles and ensuring fairness and security in a decentralized market are significant challenges.

\subsubsection{Dynamic Pricing Optimization with Real-time Contextual Understanding}

Traditional dynamic pricing algorithms often rely on historical load patterns and weather forecasts. LLMs can significantly enhance these algorithms by incorporating a wider array of real-time contextual data, including grid conditions, weather patterns, public events, and even social media trends. By understanding this rich context, an LLM can dynamically adjust energy prices to better balance supply and demand, optimize grid utilization, and incentivize desired consumer behavior, such as shifting load to off-peak hours.

LLMs enable more responsive and nuanced pricing strategies that can react to unforeseen circumstances and incentivize specific behaviors more effectively than pre-set algorithms. For example, an LLM could detect a sudden surge in demand due to a heatwave and adjust prices accordingly to prevent grid overload. Implementation requires integration with real-time data feeds from various sources including grid sensors, weather APIs, event calendars, and potentially social media monitoring tools. Training the LLM involves reinforcement learning techniques, where the LLM learns optimal pricing strategies based on the observed responses of consumers to different price signals and contextual factors. Clear communication of the dynamic pricing logic to consumers is crucial for building trust and acceptance.

\subsection{Cluster 3: Personalized Energy Management \& Customer Engagement}

This cluster focuses on how LLMs can transform the way energy is managed and how customers engage with their energy providers. The use cases here demonstrate the power of LLMs to provide personalized energy-saving recommendations, proactive outage communication, tailored tariff recommendations, and seamless smart home integration. By leveraging their natural language processing and data analysis capabilities, LLMs can create a more user-centric and responsive energy ecosystem.

\subsubsection{Personalized Energy Saving Recommendations with Conversational Coaching}

Generic energy-saving tips are often ignored as they fail to resonate with individual needs and lifestyles. LLMs can overcome this by analyzing a user's specific energy consumption patterns, appliance usage habits, lifestyle preferences, and even financial considerations. Based on this personalized understanding, LLMs can provide tailored advice through engaging conversational interfaces such as chatbots or virtual assistants. For example, instead of a generic "unplug unused electronics," an LLM might say, "Based on your usage patterns, you could save \$5 per month by unplugging your rarely used gaming console and the second refrigerator in the garage." The conversational aspect allows for follow-up questions and personalized support, making the advice more actionable.

Traditional methods rely on static websites, generic brochures, or infrequent email blasts with standardized tips, lacking personalization and adaptability. LLMs offer a dynamic and interactive experience, increasing user engagement and the likelihood of behavior change. Conversational coaching addresses user concerns, provides encouragement, and tracks progress, leading to more sustained energy savings. Implementation requires integration with smart meter data platforms and potentially with smart home device ecosystems. Developing a robust conversational interface, powered by an LLM, is crucial. Training the LLM involves incorporating behavioral science principles, energy efficiency best practices, and the ability to understand and respond to user queries and concerns. Privacy considerations regarding user data are crucial.

\subsubsection{Proactive Outage Communication \& Personalized Resolution Support}

Power outages are a major inconvenience, and lack of timely information can exacerbate customer frustration. LLMs can revolutionize outage communication by generating personalized messages to affected customers, providing real-time updates on the outage status, and offering tailored troubleshooting advice. For instance, a customer might receive a message stating, "We are aware of an outage affecting your area due to a downed power line. Crews are on-site, and the estimated restoration time is 6:00 PM. If your neighbor's power is on, please check your breaker box." This level of personalization makes the communication more relevant and helpful.

Traditional outage communication often involves generic messages through websites or social media, lacking specific information. Call centers can become overwhelmed, leading to long wait times. LLMs can automate personalized communication, reducing call center load and providing more timely and accurate information, improving satisfaction and reducing anxiety. Implementation requires tight integration with Outage Management Systems (OMS), customer databases, and potentially smart meter data. Training the LLM involves incorporating knowledge of common outage causes, troubleshooting steps, communication best practices for emergencies, and the ability to generate empathetic and informative messages. Integration with communication channels like SMS, email, and mobile apps is also necessary.

\subsubsection{Personalized Energy Tariff Recommendations with Explainable AI}

Choosing the right energy tariff can be complex, with various options based on time-of-use, demand charges, and renewable energy credits. LLMs can analyze a customer's historical energy consumption patterns, financial information, and lifestyle preferences to recommend the most cost-effective tariff. Crucially, the LLM can provide clear explanations for its recommendations, outlining potential savings and highlighting why a specific tariff is a good fit for their circumstances. For example, it might explain, "Based on your high energy usage during off-peak hours and your EV charging habits, switching to the 'Time-of-Use Saver' tariff could save you approximately \$30 per month."

Traditional tariff recommendation tools often rely on simple calculators or generic advice, which may not accurately reflect potential savings for individuals. The lack of explanation can make customers hesitant to switch. LLMs, with their ability to analyze complex data and provide clear justifications, build trust and transparency, leading to higher adoption rates and increased satisfaction. Implementation requires secure integration with smart meter data platforms, tariff information databases, and potentially customer profile data. Implementing Explainable AI (XAI) techniques is essential to ensure the LLM can provide transparent explanations. Training the LLM involves incorporating knowledge of different tariff structures, consumption patterns, and financial analysis principles.

\subsubsection{Smart Home Energy Management Integration with Natural Language Control}

Managing energy consumption across various smart home devices can be cumbersome using separate apps. LLMs can act as a central, intelligent hub, allowing users to control and optimize their energy usage through simple natural language commands. For example, a user could say, "Lower the thermostat by two degrees when everyone leaves for work," or "Charge my EV only during off-peak hours," or "Turn off all the lights in the living room." The LLM understands the intent behind these commands and translates them into specific actions for the connected devices.

Traditional smart home control often requires navigating multiple apps and interfaces, which can be inconvenient. LLMs provide a more intuitive and user-friendly experience, making it easier for users to manage their consumption effectively. This leads to greater adoption of energy-saving behaviors and increased convenience.  Implementation requires integration with the APIs of various smart home device ecosystems. Training the LLM involves mapping natural language commands to specific device functions and understanding user intentions related to energy saving. Robust security measures are crucial to protect user privacy and prevent unauthorized access to smart home devices.

\subsection{Cluster 4: Smart Grid Planning \& Education}

This cluster focuses on the role of LLMs in enhancing smart grid planning and education. The use cases here demonstrate how LLMs can automate the creation and maintenance of digital twins, generate novel grid infrastructure designs, personalize public education on smart grid technologies, and facilitate interactive training for grid operators. By leveraging their ability to process diverse data, generate creative content, and understand natural language, LLMs can significantly improve both the efficiency and effectiveness of planning and training in the energy sector.

\subsubsection{Automated Grid Digital Twin Creation \& Dynamic Updating}

Creating and maintaining an accurate digital twin of the grid is crucial for planning, simulation, and operational analysis, but manually constructing and updating these twins is labor-intensive and time-consuming. LLMs can automate this process by analyzing diverse data sources, including grid design documents, sensor readings, maintenance records, and even drone or satellite imagery. The LLM extracts relevant information, identifies relationships between components, and creates a comprehensive and up-to-date virtual representation of the grid.

Traditional methods of creating digital twins involve manual data entry and modeling, prone to errors and requiring significant effort. LLMs significantly reduce the time and cost associated with digital twin creation and maintenance, ensuring the twin accurately reflects the current grid state. This enables more accurate simulations, better informed planning decisions, and improved operational efficiency. Implementation requires robust data integration capabilities to access and process data from various sources. Training the LLM involves providing it with a vast amount of grid design documentation, engineering specifications, and data modeling principles. Developing mechanisms for the LLM to continuously update the digital twin based on real-time data is a key challenge.

\subsubsection{Generative AI for Novel Grid Infrastructure Design \& Optimization}

Designing and optimizing grid infrastructure is a complex engineering challenge involving numerous constraints. LLMs, as a form of generative AI, can explore a wider range of design possibilities than human engineers might typically consider. By feeding the LLM with design parameters, constraints, and performance objectives, it can generate novel grid layouts, substation configurations, and network topologies, potentially leading to more efficient, resilient, and cost-effective grid infrastructure.

Traditional grid design relies heavily on established engineering practices and human expertise, whereas LLMs can introduce innovative and unconventional design concepts that may not be immediately apparent. This can lead to breakthroughs in grid efficiency, reliability, and the integration of renewable energy sources. Implementation requires training LLMs on a comprehensive dataset of existing grid designs, engineering principles, material properties, cost data, and performance metrics. Defining clear constraints and objectives for the generative process is crucial. Integrating the LLM with grid simulation tools allows for evaluating the performance of the generated designs.

\subsubsection{Public Awareness \& Education on Smart Grid Technologies with Personalized Content Generation}

Public understanding and acceptance are crucial for the successful deployment of smart grid technologies. However, their technical complexities can be a barrier to widespread understanding. LLMs can help bridge this gap by creating engaging and easily understandable educational content tailored to different audiences. For example, an LLM could generate a simplified explanation of smart meters for homeowners, a more technical overview of distributed generation for energy professionals, or an interactive chatbot to answer specific questions about smart grid benefits and concerns.

Traditional public awareness campaigns often involve generic brochures or website content that may not resonate with all audiences. LLMs can personalize the educational experience, addressing specific concerns and tailoring information to the knowledge level of the individual. This can lead to greater public engagement and a better understanding of the benefits of smart grid technologies. Implementation involves training the LLM with comprehensive information on smart grid concepts, technologies, benefits, and concerns. Developing platforms for delivering personalized content is essential, such as chatbots or online educational modules. Understanding the nuances of different audiences and tailoring the language and complexity of the content is crucial.

\subsubsection{Interactive Training Simulators for Grid Operators using Natural Language Interaction}

Training grid operators to handle complex and dynamic situations is critical for ensuring grid reliability and safety. LLMs can enhance traditional training simulators by enabling natural language interaction. Instead of clicking through menus or using predefined commands, trainees can interact with the simulated grid environment using natural language. For example, a trainee could ask, "What is the current voltage at substation Alpha?" or command, "Isolate the fault on feeder Beta." The LLM interprets these commands and adjusts the simulation accordingly, providing a more realistic and intuitive training experience.

Traditional training simulators often rely on pre-programmed scenarios and limited interaction options. LLMs enable more flexible and dynamic training scenarios, allowing trainees to explore different situations and practice their decision-making skills in a more realistic environment. The natural language interface makes the training more engaging and accessible. Implementation involves integrating LLMs with existing grid simulation software. Training the LLM requires a comprehensive understanding of grid operations procedures, emergency response protocols, and grid-specific terminology. Developing mechanisms for the LLM to accurately interpret operator commands and translate them into actions within the simulation is a technical challenge.

\subsection{Cluster 5: Smart Grid Security \& Compliance}

This cluster highlights the critical role of LLMs in enhancing the security and compliance of smart grids. The use cases demonstrate how LLMs can proactively detect cybersecurity threats, automate regulatory compliance monitoring, and assess vulnerabilities within grid control systems. By leveraging their abilities in natural language processing, code analysis, and threat intelligence, LLMs can significantly improve the resilience and trustworthiness of smart grid infrastructure.

\subsubsection{Proactive Cybersecurity Threat Detection \& Natural Language Alerting}

Smart grids are increasingly vulnerable to cyberattacks due to their interconnected nature and reliance on digital technologies. Detecting sophisticated and novel threats requires advanced analysis of various data sources. LLMs can play a crucial role by analyzing network traffic patterns, system logs, threat intelligence feeds, and even security reports written in natural language. By understanding the context and semantics of this information, LLMs can identify subtle indicators of compromise (IOCs) and anomalies that might signal an ongoing or imminent cyberattack. Upon detecting a potential threat, the LLM can generate detailed, human-readable alerts explaining the nature of the threat, affected systems, and potential impact, making it easier for security analysts to respond effectively. For example, the alert might read: "Anomaly detected: Unusual login activity originating from an external IP address attempting to access critical SCADA servers in substation Alpha. This behavior matches patterns associated with the 'Energetic Bear' APT group known for targeting energy infrastructure."

Traditional cybersecurity systems often rely on signature-based or rule-based systems, which struggle with zero-day exploits or multi-stage attacks and generate a high volume of false positives. LLMs, by learning complex patterns and understanding context, can detect novel attacks and reduce false positives. Natural language alerts improve the speed and effectiveness of incident response by providing clear, actionable information. Implementation requires robust integration with cybersecurity tools and data sources, including SIEM, IDPS, firewall logs, and threat intelligence platforms. Training involves feeding the LLM massive datasets of labeled cybersecurity data, attack narratives, and expert analyses of security vulnerabilities. Developing the LLM’s ability to correlate information from different sources, understand security terminology, and generate concise alerts is a key technical challenge. Real-time processing capabilities are crucial for timely threat detection.

\subsubsection{Automated Regulatory Compliance Monitoring \& Reporting with Legal Document Analysis}

The energy sector is heavily regulated, and ensuring compliance is a significant undertaking. LLMs can automate the process of monitoring regulatory changes and generating compliance reports. By analyzing legal documents, regulatory guidelines, policy papers, and industry standards, the LLM can identify relevant obligations, track changes in regulations, and map these requirements to specific grid operations and IT systems. The LLM can then automatically generate reports demonstrating compliance, reducing manual effort and minimizing the risk of non-compliance. For instance, an LLM could analyze a new NERC CIP standard and automatically generate a report outlining specific grid assets and security procedures that need updating.

Traditional compliance monitoring involves manual review of lengthy legal documents and spreadsheets, which is time-consuming, error-prone, and difficult to scale. LLMs improve efficiency and accuracy by automating the extraction of relevant information, identification of key requirements, and generation of compliance documentation. This reduces the burden on compliance teams and minimizes the risk of fines. Implementation requires training the LLM on a comprehensive dataset of relevant legal and regulatory documents specific to the energy sector. The LLM needs to understand legal language, identify key obligations and deadlines, and map these to technical aspects of grid operations. Integration with compliance management systems and reporting tools is necessary. Ensuring the LLM's interpretation of legal documents is accurate and reliable requires careful validation by legal experts.

\subsubsection{Vulnerability Assessment \& Automated Patch Recommendation through Code Analysis}

Grid control systems rely on software and firmware, which can contain security vulnerabilities that could be exploited by attackers. Manually identifying these vulnerabilities through code review is a time-consuming and resource-intensive process. LLMs can automate this process by analyzing the source code of grid control software and firmware, comparing it against known vulnerability patterns and secure coding best practices. Upon identifying potential vulnerabilities, the LLM can recommend appropriate security patches or configuration changes. For example, an LLM might identify a buffer overflow vulnerability in a specific PLC firmware version and recommend upgrading to a patched version or implementing a specific firewall rule as a workaround.

Traditional vulnerability assessment methods often rely on static code analysis tools, which may miss subtle vulnerabilities, or manual code review, which is effective but time-consuming. LLMs can analyze code at scale, identify a wider range of vulnerabilities, and provide more context-aware patch recommendations by understanding the specific function and potential impact of the vulnerable code. Implementation requires providing the LLM with access to source code repositories. Training involves feeding the LLM a dataset of known vulnerabilities, secure coding guidelines, and examples of vulnerable and secure code. The LLM needs to understand different programming languages and identify coding errors that could lead to security flaws. Integrating the LLM with patch management systems could enable automated deployment of recommended patches after appropriate testing.

\subsection{Cluster 6: Advanced Data Analysis \& Knowledge Discovery}

This cluster explores the powerful capabilities of LLMs in advanced data analysis and knowledge discovery within smart grids. The use cases demonstrate how LLMs can generate insightful narratives from complex data, integrate cross-domain knowledge to foster innovation, and provide semantic interpretations of sensor data for better decision-making. By leveraging their ability to synthesize information from diverse sources and understand complex relationships, LLMs offer the potential to uncover hidden patterns, promote cross-disciplinary collaboration, and enhance the operational intelligence of the smart grid.

\subsubsection{Grid-Wide Narrative Generation \& Causal Relationship Discovery}

Smart grids generate massive amounts of data, but understanding the complex relationships and sequences of events that lead to specific outcomes, such as outages or equipment failures, is a significant challenge. LLMs can analyze this vast dataset, including sensor readings, event logs, operator reports, and even external data like weather patterns, to construct coherent narratives about grid events. By identifying temporal relationships, correlations, and potential causal links between seemingly unrelated incidents, LLMs can provide a deeper understanding of complex grid behavior. For example, the LLM might analyze data from multiple substations and identify a sequence of events where a weather-related surge at one substation triggered a cascading failure at several downstream locations.

Traditional data analysis techniques often focus on identifying correlations or anomalies in individual data streams. LLMs can go beyond this by synthesizing information from multiple sources, constructing a holistic picture of events, and revealing subtle causal relationships that might otherwise be missed. This deeper understanding is invaluable for troubleshooting complex issues, improving grid resilience, and preventing future incidents. Implementation requires feeding the LLM with diverse grid data streams in a unified format, robust data integration, and preprocessing pipelines. Training involves techniques for identifying temporal dependencies, event correlation, and causal inference. Developing methods for the LLM to generate human-readable and insightful narratives is a key challenge. Visualizations and interactive interfaces can enhance the utility of these narratives.

\subsubsection{Cross-Domain Knowledge Integration \& Innovation Discovery}

The smart grid domain is inherently multidisciplinary, involving expertise in power systems engineering, cybersecurity, communication networks, data analytics, and economics. Knowledge silos often exist between these domains, hindering collaboration and the discovery of innovative solutions. LLMs can act as a bridge, understanding and translating domain-specific terminology, concepts, and research findings. By analyzing research papers, patents, technical reports, and operational data from various domains, LLMs can identify connections, overlaps, and potential applications of knowledge from one domain to another, fostering cross-disciplinary innovation. For example, an LLM might identify a novel cybersecurity technique used in the telecommunications industry that could be adapted to improve the security of smart grid communication networks.

Traditionally, innovation relies on individual researchers or interdisciplinary teams manually reviewing literature and attending conferences to identify connections between different fields. LLMs can accelerate this process by automating the discovery and synthesis of knowledge across domains, potentially uncovering novel solutions and accelerating the pace of innovation in the smart grid. Implementation requires creating a comprehensive knowledge base incorporating information from various smart grid domains. This involves curating and preprocessing research papers, technical reports, patents, and standards documents. Training the LLM involves techniques for understanding domain-specific language, identifying key concepts and relationships, and summarizing information from different sources. Developing tools for visualizing these cross-domain connections and facilitating collaboration is important.

\subsubsection{Semantic Sensor Data Interpretation \& Contextual Insight Generation}

Smart grids are equipped with numerous sensors that generate vast amounts of numerical data. However, the raw sensor data often lacks context and is difficult for humans to interpret directly. LLMs can add a layer of semantic understanding by interpreting the sensor data in the context of grid topology, equipment specifications, historical patterns, and external factors like weather conditions. By understanding the "meaning" behind the sensor readings, LLMs can generate more insightful and actionable information for grid operators. For example, instead of simply reporting a voltage reading of "235 kV," the LLM might generate an insight stating, "The voltage at substation Bravo is slightly elevated (235 kV), which is unusual for this time of day and under the current load conditions, suggesting a potential issue with the voltage regulator."

Traditional sensor data analysis often relies on simple threshold-based alerts or statistical analysis, which may miss subtle anomalies or fail to provide actionable context. LLMs can provide a richer and more nuanced interpretation of sensor data, enabling operators to better understand the state of the grid and identify potential problems before they escalate. Implementation requires feeding the LLM with streams of sensor data along with relevant metadata and contextual information. Training involves linking sensor readings to real-world events, equipment behavior, and expert knowledge about normal and abnormal operating conditions. Developing methods for the LLM to generate concise, informative, and actionable insights is key.

\subsection{Cluster 7: Emerging Applications \& Societal Impact}

This cluster highlights the emerging applications of LLMs and their significant societal impact on smart grids. These use cases demonstrate how LLMs can contribute to environmental justice, optimize grid projects based on community feedback, create comprehensive asset biographies, and improve emergency response coordination across languages. By leveraging their abilities to process complex data, understand social dynamics, and bridge communication gaps, LLMs can help create more equitable, resilient, and inclusive energy systems.

\subsubsection{Environmental Justice Analysis \& Grid Impact Assessment}

Ensuring equitable access to reliable and affordable energy is a critical societal goal. LLMs can analyze the potential environmental justice implications of grid operations and planning decisions. By analyzing data on the location of polluting facilities, historical patterns of infrastructure investment, demographic data, and environmental indicators, LLMs can identify potential disparities in the distribution of environmental burdens and benefits related to the grid. For example, an LLM might analyze historical data and find that low-income communities of color are disproportionately located near older, less efficient power plants or experience more frequent and prolonged power outages. This analysis can inform policy decisions and infrastructure investments to promote environmental justice.

Traditional environmental impact assessments may not fully capture the nuances of environmental justice concerns or overlook historical patterns of inequity. LLMs can process and analyze large, complex datasets to identify subtle patterns of disparity and provide a more comprehensive understanding of the environmental justice implications of grid-related activities. Implementation requires integrating LLMs with diverse datasets, including environmental data from regulatory agencies, demographic data, GIS data on the location of grid infrastructure, and historical data on investments and outage frequency. Training the LLM involves providing information on environmental justice principles, relevant legal frameworks, and methodologies for assessing environmental burdens. Ensuring data privacy and avoiding bias is crucial.

\subsubsection{Social Impact Optimization for Grid Projects with Community Feedback Analysis}

Grid infrastructure projects can have significant social impacts on local communities. Understanding and addressing community concerns is crucial for ensuring successful project implementation and fostering positive relationships between utilities and the public. LLMs can analyze various sources of community feedback, including social media posts, news articles, public forum transcripts, and survey responses, to gauge public sentiment towards proposed grid projects, identify key concerns, and understand the potential social impacts. This information can help utilities optimize project design and implementation to minimize negative impacts and maximize community benefits. For example, an LLM might analyze social media discussions and identify community concerns about the visual impact of a proposed transmission line, leading the utility to consider undergrounding options.

Traditional methods of gathering community feedback can be time-consuming and may not capture the full range of public opinion. LLMs can analyze large volumes of unstructured data in real-time, providing a more comprehensive understanding of community sentiment and concerns. Implementation requires integrating LLMs with social media APIs, news aggregators, and platforms for collecting community feedback. Training the LLM involves natural language processing techniques for sentiment analysis, topic modeling, and identifying key stakeholders and their concerns. Ensuring data privacy and addressing potential biases in the analyzed data are important considerations.

\subsubsection{Grid Asset Biography Generation for Enhanced Lifecycle Management}

Effectively managing the lifecycle of grid assets requires a comprehensive understanding of their history, performance, and condition. LLMs can create detailed "life stories" or biographies for individual grid assets by combining information from various sources, including maintenance logs, repair records, sensor data, inspection reports, and environmental conditions. These biographies provide valuable context for decision-making regarding maintenance scheduling, equipment replacement, and risk assessment. For example, an LLM might generate a biography for a specific transformer, detailing its installation date, maintenance history, performance metrics, significant events, and predicted remaining useful life.

Traditional asset management systems often store data in a structured format, making it difficult to access and synthesize the complete history of an asset. LLMs can process unstructured data alongside structured data to create a richer and more comprehensive understanding of each asset's life story. Implementation requires integrating LLMs with asset management systems, maintenance management systems, sensor data platforms, and potentially inspection report databases. Training the LLM involves understanding different types of grid assets, their typical lifecycles, common failure modes, and relevant terminology used in maintenance and inspection reports.

\subsubsection{Cross-Lingual Grid Emergency Response Coordination}

During grid emergencies, effective communication with the public is crucial, and in diverse communities, language barriers can hinder communication and delay response efforts. LLMs can facilitate cross-lingual communication by instantly translating instructions, warnings, and information between grid operators and the public in different languages. For example, during a hurricane-related outage, an LLM could translate messages about safety precautions, estimated restoration times, and available resources into multiple languages for affected residents.

Traditional translation services can be slow and may not be readily available during emergencies. LLMs offer real-time, automated translation, ensuring that critical information reaches everyone regardless of their primary language. Implementation requires integrating LLMs with communication channels used during grid emergencies, such as public address systems, social media platforms, and emergency alert systems. Training the LLM involves providing it with a comprehensive vocabulary of grid-related terminology and emergency response phrases in multiple languages. Ensuring the accuracy and cultural appropriateness of the translations is crucial.

\subsection{Cluster 8: LLM-Enhanced Reinforcement Learning in Smart Grids}

This cluster explores the powerful synergy between LLMs and RL in smart grid applications, with a particular focus on enhancing the interaction between AI agents and human participants. The use cases highlight how LLMs can provide intuitive interfaces for controlling RL agents, generate realistic and challenging training scenarios, and, crucially, model the complex and diverse behaviors of prosumers within the grid environment. By bridging the gap between AI control and human behavior, LLMs are unlocking new possibilities for developing more effective and human-centered RL applications in smart grids.

\subsubsection{Natural Language Interface for Reinforcement Learning Agents in Grid Control}

RL is a powerful tool for optimizing complex smart grid systems, but the decision-making process of RL agents can be opaque. Instead of simply providing explanations, LLMs can serve as an intermediary, translating human requests into a format that RL agents can understand. For example, an operator might instruct the LLM: "Adjust the RL algorithm to prioritize grid stability over cost for the next hour." The LLM would then analyze the underlying code of the RL algorithm, potentially modifying its objective function or constraining its action space based on the operator's input.

Traditional interaction with RL agents often involves complex numerical manipulation and predefined parameters. LLMs can enable human operators to exert influence over RL systems in a more intuitive way, enabling more effective collaboration. Implementation involves developing a communication layer between the LLM and the RL agent. Training the LLM involves providing it with a thorough understanding of the RL agent's codebase, reward structure, and potential areas of influence. It is crucial for the LLM to translate user instructions into accurate changes to the underlying RL algorithm.

\subsubsection{LLM-Generated Scenarios for Training Robust Reinforcement Learning Agents}

Training effective RL agents for grid control requires exposing them to a wide range of realistic and challenging scenarios, including extreme events like cyberattacks or natural disasters. Manually creating these scenarios can be time-consuming and may not cover all potential contingencies. LLMs can automate the generation of diverse and realistic scenarios by analyzing historical grid data, expert knowledge about potential disruptions, and even news reports about past incidents. The LLM can generate textual descriptions of these scenarios, which can then be translated into simulation parameters for training the RL agent.

Traditional methods of creating training scenarios may be limited in scope and diversity. LLMs can generate a much wider range of scenarios, including rare or unexpected events, leading to RL agents that are more robust and adaptable to unforeseen circumstances. Implementation requires integrating LLMs with grid simulation environments. Training the LLM involves providing it with data on historical grid events, expert knowledge about potential disruptions, and the ability to generate plausible and challenging scenarios. The generated scenarios need to be translated into a format that can be used by the grid simulator to train the RL agent.

\subsubsection{LLMs for Realistic Prosumer Behavior Modeling in Reinforcement Learning}

In many smart grid applications, RL agents are being developed to interact directly with human participants, such as consumers or prosumers. Traditionally, these human actors are treated as a component of the RL agent's environment, with their collective responses forming the feedback the agent receives. Existing methods to model this human-agent interaction, like bilevel optimization or mean field game theory, often rely on simplifying assumptions. These can include the assumption that all prosumers will behave in the same way or that their decisions will always be perfectly rational. In some instances, consumer reactions are modeled as if individuals are solving complex optimization problems, which does not always reflect real-world decision-making.

LLMs offer an alternative approach. Instead of relying on these simplified mathematical models, LLMs can be trained to consider a much wider range of factors influencing prosumer behavior. This includes their varied motivations (such as cost savings, environmental concerns, or convenience), prevailing market conditions, and typical reactions to price signals and other incentives. By learning from extensive data on human behavior, LLMs can provide a more realistic representation of how prosumers will respond to the actions of an RL agent. This leads to more accurate simulations of the grid environment and the development of more effective RL agents.

Unlike traditional methods that assume uniform or perfectly rational behavior, LLMs can capture the inherent diversity and occasional unpredictability of human actions. This results in simulations that more closely mirror real-world conditions, allowing RL agents to be trained in more realistic scenarios. Furthermore, utilizing an LLM for this modeling purpose can significantly reduce the computational burden associated with running the complex simulations required by other techniques. Implementing this approach requires training the LLM on substantial datasets capturing prosumer behavior within the energy domain. This data could encompass historical energy consumption patterns, observed responses to price signals, insights from surveys and behavioral studies, and relevant market data. The trained LLM would then be integrated into the RL training loop, predicting the collective prosumer response to an RL agent's action, providing feedback for the agent's learning process. A key implementation challenge involves ensuring the LLM accurately reflects real-world behavior and avoids introducing unintended biases into the simulation. 

Table \ref{tab:llm_use_cases} provides a concise overview of the 30 LLM use cases explored in this paper, organized by cluster. For each application, the table summarizes its core motivation, the key benefits it offers over traditional methods, and the main hurdles anticipated during implementation. In addition, \ref{appendix} illustrates these use cases through practical scenarios, offering examples of how LLMs can be applied across the smart grid ecosystem.

\begin{longtable}{>{\RaggedRight\arraybackslash}p{1cm} >{\RaggedRight\arraybackslash}p{3cm} >{\RaggedRight\arraybackslash}p{3.5cm} >{\RaggedRight\arraybackslash}p{3.5cm} >{\RaggedRight\arraybackslash}p{4.5cm}}
\caption{LLM Use Cases in Smart Grids}\label{tab:llm_use_cases}\\
\toprule
\textbf{Clust.} & \textbf{Application} & \textbf{Motivation} & \textbf{Benefit} & \textbf{Challenges} \\ 
\midrule
\endfirsthead
\multicolumn{5}{c}{\tablename\ \thetable{} -- Continued from previous page} \\
\toprule
\textbf{Clust.} & \textbf{Application} & \textbf{Motivation} & \textbf{Benefit} & \textbf{Challenges} \\ 
\midrule
\endhead
\midrule
\multicolumn{5}{r}{Continued on next page} \\
\endfoot
\bottomrule
\endlastfoot
\textbf{1} & Reconfig \& Opt & Complex grids, manual hard & Agile, intuitive, faster & Data pipeline, safety, reliability \\
\midrule
\textbf{1} & Anomaly Detect & Data overload, real threats hidden & Reduce false positives, context & Data integration, explanation accuracy \\
\midrule
\textbf{1} & Resilience Assess & Complex interdependencies, cascades  & Detailed scenarios, better visualization & Data integration, plausible narratives \\
\midrule
\textbf{1} & Fault Loc \& Repair & Minimize downtime, quick repair & Faster, automate troubleshooting & Data integration, visual info \\
\midrule
\textbf{1} & Microgrid Manage & Opt DERs, ensure stability & Dynamic, context aware & Robust communication \\
\midrule
\textbf{2} & PPA Negot & Complex, legal heavy, time consume & Streamlined, transparent, optimal terms & Legal, ethical adherence \\
\midrule
\textbf{2} & Market Sentiment & Influenced by many factors & Nuanced, real-time info & Access to diverse data \\
\midrule
\textbf{2} & Decentral Trade & Centralized markets, small players & Better prices, localized system & Secure platform, regulatory hurdles \\
\midrule
\textbf{2} & Dyn Price Opt &  Static pricing algorithms & Responsive to context, prevent overload & Real-time data, consumer trust \\
\midrule
\textbf{3} & Pers. Saving & Generic tips not engaging & Dynamic, interactive, behavior change & Smart meter data, privacy \\
\midrule
\textbf{3} & Outage Comm & Lack of info, frustration & Personalized, timely, accurate & Outage systems, empathetic messages \\
\midrule
\textbf{3} & Tariff Recom & Complex, hard to choose right plan & Clear, understandable explanation & Detailed consumption history \\
\midrule
\textbf{3} & Smart Home Int & Multiple apps, cumbersome & Intuitive, user friendly & API integration, device security \\
\midrule
\textbf{4} & Digital Twin & Manual process, time-consuming & Reduces time, accurate twin & Diverse data, continuous update \\
\midrule
\textbf{4} & Novel Design & Complex engineering, constraints & Innovative, unconventional designs & Comprehensive data set \\
\midrule
\textbf{4} & Public Aware & Technical complex, low understanding & Personalized, engaging content & Tailored content delivery \\
\midrule
\textbf{4} & Train Simulators & Predefined scenarios, limited options & Flexible, dynamic training &  Interpret commands, translate to sim \\
\midrule
\textbf{5} & Threat Detect & Interconnect, reliance on digital & Detect novel attacks, reduce false alerts & Real-time data, security terminology \\
\midrule
\textbf{5} & Compliance Monit & Heavily regulated sector & Automate reporting, improve efficiency & Understand legal language, map req \\
\midrule
\textbf{5} & Vulnerability Assess & Software vulnerabilities, time consuming & Analyze code, automate patching & Source code access, code analysis \\
\midrule
\textbf{6} & Grid Narrative & Complex relationships, vast data  & Holistic event, reveal causal links & Diverse data, human-readable narratives \\
\midrule
\textbf{6} & Cross-Domain Knowledge & Knowledge silos & Accelerate innovation, bridge the gap & Comprehensive knowledge, visualization \\
\midrule
\textbf{6} & Semantic Sensor & Raw data lacks context &  Actionable insights for operator & Link readings to events, insights generation \\
\midrule
\textbf{7} & Environ. Justice  &  Equity, affordability goals & Identify disparities, better decisions & Diverse data, avoid bias \\
\midrule
\textbf{7} & Social Impact & Public opinion, local communities & Nuance, comprehensive, real-time & Integrate data, address biases \\
\midrule
\textbf{7} & Asset Biography & Full asset history & Synthesize complete history of an asset & Integrate systems and historical data \\
\midrule
\textbf{7} & Cross-Lingual Res. & Language barrier, delays & Real time, automated translations & Multiple languages, cultural appropriateness \\
\midrule
\textbf{8} & RL Interface & RL agents as black boxes & Translates human requests to RL & Communication layer, modify RL settings, ensure accuracy \\
\midrule
\textbf{8} & RL Scenarios & Limited scope, not diverse & Robust and adaptable RL agents & Integrate data, generate plausible scenarios \\
\midrule
\textbf{8} & Realistic Prosumer Modeling & Capturing diverse human behavior in RL & More accurate simulations, reduced computational cost & Training data on prosumer behavior, avoiding bias \\
\bottomrule
\end{longtable}

\section{Technical Challenges and Solutions}
\label{sec:challenges}

Integrating LLMs into smart grids, while promising transformative advancements, necessitates careful consideration of several interconnected technical challenges. To fully realize the potential of LLMs in this domain, we must effectively address key hurdles spanning data privacy and security, computational resource management, system interoperability, model reliability and trustworthiness, scalability, explainability, and regulatory compliance.

Firstly, data privacy and security emerge as paramount concerns. Smart grids inherently process sensitive information, including granular consumption patterns and critical operational data. Integrating LLMs can amplify these risks, potentially exposing private behaviors \citep{Escobar2021} or proprietary operational details. Fortunately, research offers viable solutions. Federated learning \citep{Li2022} enables LLM training on decentralized data while preserving data origin privacy. Differential privacy adds calibrated noise to datasets, preventing individual re-identification. Secure Multi-Party Computation \citep{Rottondi2013} allows computation on encrypted data, and secure enclaves provide isolated environments for sensitive data processing. Beyond these techniques, robust data protection relies on essential measures like anonymization, role-based access controls, strong encryption, and regular security audits \citep{Munshi2018}. Furthermore, blockchain technology offers a promising avenue for enhanced data privacy, particularly in decentralized energy trading \citep{Alladi2019}.

In addition, recent work highlights novel cybersecurity risks specific to LLMs, such as false or malicious data injection and unauthorized information extraction \citep{li_risks_2024}. These align with prompt-injection and data poisoning attacks, potentially allowing adversaries to manipulate or extract critical grid knowledge. Guardrail mechanisms, including layered filtering of inputs and adversarial training, can mitigate these LLM-specific threats\citep{ayyamperumal_current_2024}.

Beyond data protection, managing the substantial computational resources demanded by LLMs is a crucial challenge. Efficient resource utilization is key for practical smart grid deployment. Techniques like quantization, pruning, and knowledge distillation \citep{Bhattarai2019} effectively reduce the model's resource footprint. Edge computing strategically moves processing closer to data sources, reducing latency and central infrastructure load. Specialized hardware, such as GPUs, TPUs, and FPGAs, can significantly accelerate computations. Dynamic resource allocation, especially in cloud-based demand-side management \citep{Cao2017}, optimizes resource use in real-time. Cost-effective deployment can be achieved through cloud and hybrid architectures, potentially using fine-tuned edge models. Additionally, incremental and transfer learning can minimize training time and computational demands.

Another layer of complexity arises from system interoperability within heterogeneous smart grids. Diverse technologies and communication protocols necessitate standardized solutions for seamless LLM integration. Standardized APIs are crucial for facilitating communication between components \citep{Zeid2019}. Middleware platforms can translate data and protocols, bridging system differences. A modular design enables gradual integration and simplifies maintenance. Semantic interoperability, enhanced by ontologies, and data wrappers for format compatibility improve system understanding. Ultimately, industry-wide standards are essential for streamlined communication and integration \citep{Cintuglu2018}.

For safety-critical smart grid operations, ensuring reliability and trustworthiness of LLM outputs is paramount. Comprehensive testing using historical and real-time simulations is essential to validate performance across diverse scenarios. To address the opacity of LLM decisions, XAI methods can foster user trust. Human-in-the-loop systems provide a vital safety mechanism through expert review. Continuous monitoring detects performance degradation, triggering retraining. Domain-specific fine-tuning enhances accuracy for specific applications \citep{Sun2019}. Furthermore, robustness against adversarial attacks and formal verification methods \citep{Farraj2018} are crucial for system safety. Research into Recurrent Neural Networks (RNNs) is also advancing anomaly detection and reliability \citep{Ayad2018}.

Considering the escalating data and computational loads in smart grids, scalability becomes a critical factor. Distributed processing architectures are key to distributing workloads and achieving scalability. Optimized data pipelines are essential for managing high-velocity data streams efficiently \citep{Munshi2018}. Load balancing and failover mechanisms ensure operational continuity. Model and data parallelism accelerate processing, while asynchronous processing enhances responsiveness.

Maintaining consistent LLM performance in dynamic smart grid environments requires continuous adaptation. Continuous learning pipelines enable regular retraining with new data, allowing LLMs to adapt to evolving conditions. Monitoring systems detect performance deviations, prompting retraining when needed. Adaptive or online learning allows incremental adjustments, minimizing full retraining cycles and ensuring models remain current.

Addressing the "black box" nature of some LLMs is crucial for user trust and acceptance, especially in operational contexts. Providing context-rich explanations for LLM decisions is therefore vital. Human-in-the-loop validation by experts adds crucial assurance. Comprehensive documentation detailing LLM behavior enhances transparency and user understanding.

Finally, successful LLM integration necessitates compliance with reliability, cybersecurity, and data protection standards. Regulatory sandboxes offer controlled environments for initial evaluation. Proactive engagement with regulatory bodies ensures adherence to evolving guidelines. Robust documentation with clear audit trails is vital for demonstrating ongoing compliance.

\section{Evaluation and Performance Metrics}
\label{sec:evals}
Successfully integrating LLMs into smart grids depends on thorough evaluation. Moving beyond basic functionality, clear performance metrics are essential for measuring benefits, ensuring reliability, and building trust. As \citet{Farzaneh2021} notes, a strong evaluation framework is key for AI in energy, applying directly to LLMs in smart grids. This section outlines key metrics for assessing LLM performance, highlighting that effective evaluation includes technical accuracy, user experience, economic impact, explainability, and user trust. The choice of metrics should align with each application's specific goals, reflecting the diverse uses of LLMs in smart grids.

Evaluation metrics can be categorized for clarity. Technical performance metrics are crucial, focusing on the LLM's ability to perform tasks accurately and efficiently. These provide a quantitative assessment of core functions. Accuracy, a fundamental metric, measures the correctness of outputs. Its measurement varies by task; for classification like anomaly detection, precision, recall, and F1-score are vital. For information retrieval, accuracy measures the LLM's ability to find relevant information. In numerical prediction, like energy demand forecasting, Mean Absolute Error (MAE), Mean Squared Error (MSE), and Root Mean Squared Error (RMSE) quantify prediction deviations, echoing the need for reliable prediction in electricity load estimation \citep{Solyali2020}. Response time (latency) is crucial for real-time applications. Throughput measures the system's data processing capacity. Robustness, assessing resilience to noise or attacks, is vital for reliability, a concern in smart grid security \citep{Lin2022, Nourani2019}. Resource utilization, like CPU and memory usage, is important for edge deployments.

User-centric metrics are also important, as LLM success depends on user acceptance and workflow integration. User satisfaction, measured by surveys, assesses overall experience. Task completion rate reflects usability, and task completion time measures efficiency gains. Learnability assesses ease of use for new users. The importance of user experience is supported by \citet{Wang2020}, who emphasize a comprehensive approach to evaluating AI in power management.

Economic metrics show the tangible value of LLMs. Cost-effectiveness balances implementation costs with benefits like reduced expenses. Return on Investment (ROI) measures profitability. Time savings quantify efficiency gains. Operational efficiency includes improved energy efficiency and reduced outage duration (MTTR). Scalability cost assesses the expense of scaling the system.

Explainability and user trust are critical for adopting LLMs in critical infrastructure. Explainability addresses the "black box" nature of some LLMs, evaluating their ability to justify decisions, as highlighted by \citet{Kuzlu2020} for solar forecasting. Articulating reasoning is key for building confidence. User trust, based on reliability and transparency, can be assessed through surveys. Factors like consistency, accuracy, robustness, and explainability influence trust. Building trust requires systems that are understandable and beneficial to users. Evaluating LLMs must include these human-centered aspects.

In conclusion, evaluating LLMs in smart grids requires a holistic approach covering technical capabilities, user experience, economic impact, and trust. A comprehensive framework should include quantitative data from technical and economic metrics and qualitative feedback from user assessments. Continuous monitoring and refinement are essential to ensure LLMs are effective, accepted, and trusted partners in creating smart, resilient, and sustainable energy systems. The focus should be on developing powerful, transparent tools that serve the needs of the grid and its users. Table \ref{tab:evaluation_metrics_summary} summarizes the key evaluation metrics discussed above.

\begin{longtable}{>{\RaggedRight\arraybackslash}p{3cm} >{\RaggedRight\arraybackslash}p{3.5cm} >{\RaggedRight\arraybackslash}p{8cm}}
\caption{Summary of Evaluation Metrics}\label{tab:evaluation_metrics_summary}\\
\toprule
\textbf{Category} & \textbf{Metric} & \textbf{Description} \\
\midrule
\endfirsthead
\multicolumn{3}{c}{\tablename\ \thetable{} -- Continued from previous page} \\
\toprule
\textbf{Category} & \textbf{Metric} & \textbf{Description} \\
\midrule
\endhead
\midrule
\multicolumn{3}{r}{Continued on next page} \\
\endfoot
\bottomrule
\endlastfoot
\multirow{5}{3cm}{\textbf{Technical}} & \textbf{Accuracy} & Correctness of LLM outputs; measured by precision, recall, F1-score (classification), MAE, MSE, RMSE (prediction) \\
\cline{2-3}
& \textbf{Response Time (Latency)} & Time taken by LLM to process input and generate a response \\
\cline{2-3}
& \textbf{Throughput} & Number of requests handled within a timeframe \\
\cline{2-3}
& \textbf{Robustness} & Resilience to noisy, incomplete, adversarial inputs \\
\cline{2-3}
& \textbf{Resource Util.} & CPU, memory, energy consumption \\
\midrule
\multirow{4}{3cm}{\textbf{User-Centric}} & \textbf{User Satisfaction} & User perception of system's usability, usefulness, reliability, trust \\
\cline{2-3}
& \textbf{Task Compl. Rate} & Proportion of times users complete intended task successfully \\
\cline{2-3}
& \textbf{Task Compl. Time} & Time to complete tasks with LLM vs traditional methods \\
\cline{2-3}
& \textbf{Learnability} & Ease with which new users can understand and utilize the LLM \\
\midrule
\multirow{5}{3cm}{\textbf{Economic}} & \textbf{Cost-Effectiveness} & Balance of implementation costs and benefits achieved \\
\cline{2-3}
& \textbf{ROI} & Profitability of LLM implementation \\
\cline{2-3}
& \textbf{Time Savings} & Reduction in task completion time using LLMs \\
\cline{2-3}
& \textbf{Operational Eff.} & Improvement in energy efficiency and outage times \\
\cline{2-3}
& \textbf{Scalability Cost} & Costs of scaling up the LLM system \\
\midrule
\multirow{2}{3cm}{\textbf{Trust \& Explainability}} & \textbf{Explainability} & Clarity of reasoning behind LLM decisions \\
\cline{2-3}
& \textbf{User Trust} & User confidence in system's reliability and transparency \\
\bottomrule
\end{longtable}

\section{Societal and Policy Implications}
\label{sec:soc}
Integrating LLMs into smart grids is a major shift with significant societal and policy implications. Beyond technical gains, we must consider their broader impact on fairness, ethics, regulation, and public interaction with energy. AI and machine learning are already improving energy management \citep{Kushawaha2024, Olatunde2024, Maurya2024}, and LLMs can enhance this further with advanced data analysis and real-time decision-making in our dynamic energy systems.

LLMs offer a powerful way to address long-standing inequalities in energy access and affordability. By analyzing detailed data, we can identify underserved communities for targeted improvements. This data-driven approach can lead to fairer resource allocation, guiding investments in better grids, renewable energy, and efficiency programs where needed most. LLMs can also provide personalized tools to help vulnerable populations manage energy and lower costs. This is supported by LLMs' ability to analyze consumption patterns, predict demand, and enable efficient energy distribution \citep{Maurya2024, Marques2024}. Policies are needed to encourage LLM use to combat energy poverty and improve grid reliability in these areas, while preventing biased pricing or resource allocation.

However, using LLMs in critical infrastructure raises ethical questions. A key concern is algorithmic bias from training data reflecting societal inequalities, potentially leading to unfair pricing or resource distribution. Addressing this requires diverse training data, continuous bias monitoring, and clear explanations for AI decisions using XAI to build trust. Furthermore, the risk of LLMs being misused for cyberattacks necessitates strong security measures \citep{Mohammed2024, Bouramdane2023} and policies promoting ethical AI and responsible data handling. Accountability is also crucial. If AI-driven control leads to an outage or financial harm, guidelines should clarify who is liable, ensuring grid operators, developers, and regulators share responsibility.

Integrating LLMs also presents regulatory challenges. Current regulations may not suit the capabilities and risks of LLMs. Adaptable policies are needed to guide LLM development and use, including clear rules for data privacy and security \citep{Mohammed2024, Bouramdane2023}. Ensuring LLM-driven controls are reliable and safe and defining liability for failures are also critical. Regulators must collaborate with stakeholders to create guidelines that encourage innovation while protecting public interests, focusing on safety approvals, consumer rights regarding dynamic pricing, and system compatibility \citep{Manas2015}. 
Regulatory agencies must mandate rigorous testing procedures or official certifications for AI systems managing essential grid functions, emphasizing human oversight to prevent unchecked automation.

LLMs can significantly improve public engagement and understanding of smart grids. They can explain complex operations and offer personalized energy-saving advice. LLM-powered chatbots can answer questions, provide tailored information, and foster trust in data use and decision-making \citep{Chung2023}. Analyzing public feedback through LLMs can provide real-time insights into community concerns, leading to more inclusive decisions and aligning infrastructure projects with local needs and environmental goals. LLMs can empower individuals to manage energy better, promoting conservation and public acceptance \citep{Ezeigweneme2024}.

Finally, LLM-driven automation will change the energy workforce. While some jobs may be automated, new roles requiring AI and data science skills will emerge. Proactive policies are needed to support affected workers through retraining, ensuring a fair transition. Investing in education to equip workers with LLM system skills is essential. Collaboration between AI specialists and power engineers is crucial for effective and reliable systems, emphasizing the importance of interdisciplinary training for improved grid operation and security \citep{Shi2024}. In conclusion, the societal and policy implications of LLMs in smart grids are wide-ranging and need careful consideration. By promoting responsible innovation, ensuring equal access, creating strong regulations, and prioritizing public involvement, we can use LLMs to build a smarter, more sustainable, and fairer energy future.

\section{Future Directions}
\label{sec:future}
Integrating LLMs into smart grids is just the beginning. Many future developments promise to significantly increase their impact and address emerging challenges in the energy sector. Advances in both AI and smart grid technologies will further enhance LLM capabilities for intelligent energy management.

One exciting future direction is the development of multi-modal LLMs. Currently, LLMs mostly use text data. Future LLMs will likely use various data types, including visual information from drones and satellites, offering insights into infrastructure conditions. They will also incorporate sensor data for real-time grid status and audio data from maintenance and field communications, creating a more complete understanding. Imagine LLMs analyzing a technician's report with images of damaged equipment to suggest repairs or combining sensor readings with event descriptions for better grid operation insights, leading to faster problem solving and better responses.

Another promising area is combining LLMs with quantum computing. Quantum computing can solve complex optimization problems that are too difficult for today's computers. Using LLMs with quantum algorithms could transform dynamic grid optimization by quickly analyzing large datasets to find the best grid configurations. This would make grids more efficient and resilient, allowing for real-time adjustments to balance energy supply and demand, especially with renewable energy sources. Quantum algorithms could also improve energy market simulations, leading to better pricing and resource allocation. LLMs could prepare data for and interpret the results of quantum computations, accelerating the discovery of new materials for better energy storage and transmission.

Future LLMs need to learn and adapt in the ever-changing smart grid environment. As new technologies are added, consumption patterns shift, and environmental conditions change, LLMs must continuously learn from new data and feedback. This continuous learning is essential for maintaining accuracy and relevance, allowing them to effectively respond to unexpected events or sudden changes in energy demand. For example, an adaptive LLM could automatically adjust its settings based on changing weather or energy consumption trends.

The future will also see the development of real-time LLM applications for active grid operations. This includes processing sensor data for immediate anomaly detection, enabling quicker responses to equipment failures or cyberattacks. LLMs can also facilitate dynamic pricing adjustments based on real-time grid conditions to balance supply and demand. Furthermore, LLMs can intelligently coordinate distributed energy resources like solar panels and batteries, optimizing energy flow and ensuring grid stability. Achieving these real-time applications requires improvements in model efficiency and the use of edge computing and specialized hardware for fast responses. In addition, deploying modular LLM architectures capable of updating selective components may allow the system to adapt quickly without retraining an entire model whenever local conditions evolve.

As LLMs become more critical to grid operations, strong validation and testing methods are essential. Future research must focus on developing ways to assess their reliability, safety, and security. This includes creating realistic simulations of various scenarios, from extreme weather to cyberattacks. Testing against malicious inputs will be crucial, as will continuous monitoring for errors, biases, or unexpected behavior. Ensuring that LLM decision-making is understandable is also vital for building trust among grid operators. These validation efforts should verify that rapid or online model updates preserve overall stability, preventing oscillations or performance regressions.

Beyond grid design, generative AI offers additional benefits. It can create synthetic datasets to improve the training of machine learning models, especially for rare events. Generative AI can also develop realistic scenarios for training simulations, improving preparedness for emergencies. Moreover, these models can create customized educational materials to better inform the public about smart grid technologies.

Focusing on human-centered design is key to fully utilizing LLMs in smart grids. These powerful tools must be easy to use, understand, and integrate into existing workflows. This requires developing user interfaces tailored to the needs of grid operators, consumers, and other stakeholders. Prioritizing user experience will bridge the gap between AI capabilities and practical use, making these tools accessible and effective for everyone. Future frameworks might include interactive dashboards where humans can fine-tune LLM behavior on the fly, maintaining operator confidence and oversight.

Finally, as LLMs become more common, addressing their ethical and societal implications is crucial. Ongoing research must examine potential biases in LLM algorithms to ensure fair outcomes. Analyzing the impact of AI-driven automation on jobs and creating strategies for a just transition is necessary. Furthermore, we must prevent the misuse of LLMs by promoting responsible deployment. This research will guide the development of policies to ensure LLMs are used to benefit society. Further research into LLMs and reinforcement learning could involve LLMs automatically creating complex training environments for RL agents and helping to understand their decisions, leading to more advanced autonomous grid control.

\section{Conclusion}
\label{sec:conclusion}
This paper has explored the vast potential of LLMs to revolutionize smart grid operations. LLMs are not just theoretical concepts; they offer practical solutions for improving grid management and empowering users. Our analysis of thirty use cases across eight areas shows LLMs' versatility in addressing complex energy challenges.

LLMs offer unique advantages by understanding natural language, reasoning through complex data, and providing relevant outputs. This enables smarter grid operations with automated fault detection and optimized resource use. In energy markets, LLMs can predict trading opportunities and support decentralized platforms. For consumers, they provide personalized energy advice, encouraging participation and conservation. LLMs also assist in planning by creating digital models and educational content, and enhance security by detecting cyber threats and ensuring regulatory compliance. Furthermore, LLMs can extract valuable insights from large datasets and support societal goals like fair energy distribution and environmental justice.

The development of LLM-enhanced reinforcement learning indicates a future with more adaptable and efficient AI agents in smart grids. This allows for advanced control strategies but also emphasizes the need for strong safety and reliability measures. While the possibilities are significant, implementing LLMs faces challenges, including data privacy, computational costs, integration with existing infrastructure, and ensuring model reliability. Solutions like federated learning, secure data handling, and thorough testing are crucial for overcoming these challenges and successfully deploying LLMs.

The main conclusion is that LLMs are a major shift in how we approach smart grid management and design, not just minor improvements. They promise more efficient, resilient, and user-focused energy systems. However, ongoing research is essential to evaluate these solutions, maintain performance and security, and minimize risks. Understanding the societal and policy impacts is also vital to ensure these technological advancements benefit everyone, particularly in terms of environmental and social justice.

Ultimately, successfully integrating LLMs into smart grids depends on continuous research and collaboration among AI specialists, energy experts, and policymakers. This combined effort is necessary to fully utilize the power of LLMs and achieve their transformative potential for the future of energy. By working together, we can create smarter, more sustainable, and equitable energy systems for all.

\section{Declaration of generative AI and AI-assisted technologies in the writing process}

During the preparation of this work, the authors used ChatGPT (developed by OpenAI) and Claude (developed by Anthropic) to improve language clarity, readability, and conciseness. After using these tools, the authors carefully reviewed and edited the content as needed, and take full responsibility for the accuracy and integrity of the published article.


\bibliography{cas-refs}

\begin{thebibliography}{79}
\expandafter\ifx\csname natexlab\endcsname\relax\def\natexlab#1{#1}\fi
\providecommand{\url}[1]{\texttt{#1}}
\providecommand{\href}[2]{#2}
\providecommand{\path}[1]{#1}
\providecommand{\DOIprefix}{doi:}
\providecommand{\ArXivprefix}{arXiv:}
\providecommand{\URLprefix}{URL: }
\providecommand{\Pubmedprefix}{pmid:}
\providecommand{\doi}[1]{\href{http://dx.doi.org/#1}{\path{#1}}}
\providecommand{\Pubmed}[1]{\href{pmid:#1}{\path{#1}}}
\providecommand{\bibinfo}[2]{#2}
\ifx\xfnm\relax \def\xfnm[#1]{\unskip,\space#1}\fi
\bibitem[{Bayindir et~al.(2016)Bayindir, Colak, Fulli, and Demirtas}]{bayindir2016smart}
\bibinfo{author}{R.~Bayindir}, \bibinfo{author}{I.~Colak}, \bibinfo{author}{G.~Fulli}, \bibinfo{author}{K.~Demirtas},
\newblock \bibinfo{title}{Smart grid technologies and applications},
\newblock \bibinfo{journal}{Renewable and sustainable energy reviews} \bibinfo{volume}{66} (\bibinfo{year}{2016}) \bibinfo{pages}{499--516}.
\bibitem[{Koliba et~al.(2014)Koliba, DeMenno, Brune, and Zia}]{koliba2014salience}
\bibinfo{author}{C.~Koliba}, \bibinfo{author}{M.~DeMenno}, \bibinfo{author}{N.~Brune}, \bibinfo{author}{A.~Zia},
\newblock \bibinfo{title}{The salience and complexity of building, regulating, and governing the smart grid: Lessons from a statewide public--private partnership},
\newblock \bibinfo{journal}{Energy Policy} \bibinfo{volume}{74} (\bibinfo{year}{2014}) \bibinfo{pages}{243--252}.
\bibitem[{Staffell and Pfenninger(2018)}]{staffell2018increasing}
\bibinfo{author}{I.~Staffell}, \bibinfo{author}{S.~Pfenninger},
\newblock \bibinfo{title}{The increasing impact of weather on electricity supply and demand},
\newblock \bibinfo{journal}{Energy} \bibinfo{volume}{145} (\bibinfo{year}{2018}) \bibinfo{pages}{65--78}.
\bibitem[{Torriti(2017)}]{torriti2017understanding}
\bibinfo{author}{J.~Torriti},
\newblock \bibinfo{title}{Understanding the timing of energy demand through time use data: Time of the day dependence of social practices},
\newblock \bibinfo{journal}{Energy research \& social science} \bibinfo{volume}{25} (\bibinfo{year}{2017}) \bibinfo{pages}{37--47}.
\bibitem[{Rodriguez-Calvo et~al.(2018)Rodriguez-Calvo, Cossent, and Fr{\'\i}as}]{rodriguez2018scalability}
\bibinfo{author}{A.~Rodriguez-Calvo}, \bibinfo{author}{R.~Cossent}, \bibinfo{author}{P.~Fr{\'\i}as},
\newblock \bibinfo{title}{Scalability and replicability analysis of large-scale smart grid implementations: Approaches and proposals in europe},
\newblock \bibinfo{journal}{Renewable and Sustainable Energy Reviews} \bibinfo{volume}{93} (\bibinfo{year}{2018}) \bibinfo{pages}{1--15}.
\bibitem[{Bekara(2014)}]{bekara2014security}
\bibinfo{author}{C.~Bekara},
\newblock \bibinfo{title}{Security issues and challenges for the iot-based smart grid},
\newblock \bibinfo{journal}{Procedia Computer Science} \bibinfo{volume}{34} (\bibinfo{year}{2014}) \bibinfo{pages}{532--537}.
\bibitem[{Chen et~al.(2025)Chen, Zhong, Long, Tang, Wang, and Liu}]{chen_advancing_2025}
\bibinfo{author}{Z.~Chen}, \bibinfo{author}{R.~Zhong}, \bibinfo{author}{W.~Long}, \bibinfo{author}{H.~Tang}, \bibinfo{author}{A.~Wang}, \bibinfo{author}{Z.~e.~a. Liu},
\newblock \bibinfo{title}{Advancing oil and gas emissions assessment through large language model data extraction},
\newblock \bibinfo{journal}{Energy and AI} \bibinfo{volume}{20} (\bibinfo{year}{2025}) \bibinfo{pages}{100481}. \URLprefix \url{https://www.sciencedirect.com/science/article/pii/S2666546825000138}. \DOIprefix\doi{10.1016/j.egyai.2025.100481}, \bibinfo{note}{accessed: 2025-02-19}.
\bibitem[{Han et~al.(2024)Han, Cong, Yu, Tang, and Wei}]{han_integrating_2024}
\bibinfo{author}{T.~Han}, \bibinfo{author}{R.-G. Cong}, \bibinfo{author}{B.~Yu}, \bibinfo{author}{B.~Tang}, \bibinfo{author}{Y.-M. Wei},
\newblock \bibinfo{title}{Integrating local knowledge with {ChatGPT}-like large-scale language models for enhanced societal comprehension of carbon neutrality},
\newblock \bibinfo{journal}{Energy and AI} \bibinfo{volume}{18} (\bibinfo{year}{2024}) \bibinfo{pages}{100440}. \URLprefix \url{https://www.sciencedirect.com/science/article/pii/S266654682400106X}. \DOIprefix\doi{10.1016/j.egyai.2024.100440}, \bibinfo{note}{accessed: 2025-02-19}.
\bibitem[{Andoni et~al.(2019)Andoni, Robu, Flynn, Abram, Geach, and Jenkins}]{andoni2019blockchain}
\bibinfo{author}{M.~Andoni}, \bibinfo{author}{V.~Robu}, \bibinfo{author}{D.~Flynn}, \bibinfo{author}{S.~Abram}, \bibinfo{author}{D.~Geach}, \bibinfo{author}{D.~e.~a. Jenkins},
\newblock \bibinfo{title}{Blockchain technology in the energy sector: A systematic review of challenges and opportunities},
\newblock \bibinfo{journal}{Renewable and sustainable energy reviews} \bibinfo{volume}{100} (\bibinfo{year}{2019}) \bibinfo{pages}{143--174}.
\bibitem[{Massaoudi et~al.(2021)Massaoudi, Abu-Rub, Refaat, Chihi, and Oueslati}]{massaoudi2021deep}
\bibinfo{author}{M.~Massaoudi}, \bibinfo{author}{H.~Abu-Rub}, \bibinfo{author}{S.~S. Refaat}, \bibinfo{author}{I.~Chihi}, \bibinfo{author}{F.~S. Oueslati},
\newblock \bibinfo{title}{Deep learning in smart grid technology: A review of recent advancements and future prospects},
\newblock \bibinfo{journal}{IEEE Access} \bibinfo{volume}{9} (\bibinfo{year}{2021}) \bibinfo{pages}{54558--54578}.
\bibitem[{Omitaomu and Niu(2021)}]{omitaomu2021artificial}
\bibinfo{author}{O.~A. Omitaomu}, \bibinfo{author}{H.~Niu},
\newblock \bibinfo{title}{Artificial intelligence techniques in smart grid: A survey},
\newblock \bibinfo{journal}{Smart Cities} \bibinfo{volume}{4} (\bibinfo{year}{2021}) \bibinfo{pages}{548--568}.
\bibitem[{Göpfert et~al.(2024)Göpfert, Weinand, Kuckertz, and Stolten}]{gopfert_opportunities_2024}
\bibinfo{author}{J.~Göpfert}, \bibinfo{author}{J.~M. Weinand}, \bibinfo{author}{P.~Kuckertz}, \bibinfo{author}{D.~Stolten},
\newblock \bibinfo{title}{Opportunities for large language models and discourse in engineering design},
\newblock \bibinfo{journal}{Energy and AI} \bibinfo{volume}{17} (\bibinfo{year}{2024}) \bibinfo{pages}{100383}. \URLprefix \url{https://www.sciencedirect.com/science/article/pii/S2666546824000491}. \DOIprefix\doi{10.1016/j.egyai.2024.100383}, \bibinfo{note}{accessed: 2025-02-19}.
\bibitem[{Omitaomu and Niu(2021)}]{Omitaomu2021}
\bibinfo{author}{O.~A. Omitaomu}, \bibinfo{author}{H.~Niu},
\newblock \bibinfo{title}{Artificial intelligence techniques in smart grid: A survey},
\newblock \bibinfo{journal}{Smart Cities 2021, Vol. 4, Pages 548-568} \bibinfo{volume}{4} (\bibinfo{year}{2021}) \bibinfo{pages}{548--568}. \URLprefix \url{https://www.mdpi.com/2624-6511/4/2/29/htm https://www.mdpi.com/2624-6511/4/2/29}. \DOIprefix\doi{10.3390/SMARTCITIES4020029}, \bibinfo{note}{accessed: 2025-02-19}.
\bibitem[{Casagrande et~al.(2014)Casagrande, Woldeamlak, Woon, Zeineldin, and Svetinovic}]{Casagrande2014}
\bibinfo{author}{E.~Casagrande}, \bibinfo{author}{S.~Woldeamlak}, \bibinfo{author}{W.~L. Woon}, \bibinfo{author}{H.~H. Zeineldin}, \bibinfo{author}{D.~Svetinovic},
\newblock \bibinfo{title}{Nlp-kaos for systems goal elicitation: Smart metering system case study},
\newblock \bibinfo{journal}{IEEE Transactions on Software Engineering} \bibinfo{volume}{40} (\bibinfo{year}{2014}) \bibinfo{pages}{941--956}. \DOIprefix\doi{10.1109/TSE.2014.2339811}.
\bibitem[{Dong et~al.(2021)Dong, Guan, Wu, Du, and Guizani}]{Dong2021}
\bibinfo{author}{J.~Dong}, \bibinfo{author}{Z.~Guan}, \bibinfo{author}{L.~Wu}, \bibinfo{author}{X.~Du}, \bibinfo{author}{M.~Guizani},
\newblock \bibinfo{title}{A sentence-level text adversarial attack algorithm against iiot based smart grid},
\newblock \bibinfo{journal}{Computer Networks} \bibinfo{volume}{190} (\bibinfo{year}{2021}) \bibinfo{pages}{107956}. \DOIprefix\doi{10.1016/J.COMNET.2021.107956}.
\bibitem[{Jha(2023)}]{Jha2023}
\bibinfo{author}{R.~K. Jha},
\newblock \bibinfo{title}{Strengthening smart grid cybersecurity: An in-depth investigation into the fusion of machine learning and natural language processing},
\newblock \bibinfo{journal}{Journal of Trends in Computer Science and Smart Technology} \bibinfo{volume}{5} (\bibinfo{year}{2023}) \bibinfo{pages}{284--301}. \DOIprefix\doi{10.36548/JTCSST.2023.3.005}.
\bibitem[{Vaswani et~al.(2017)Vaswani, Shazeer, Parmar, Uszkoreit, Jones, and Gomez}]{Vaswani2017}
\bibinfo{author}{A.~Vaswani}, \bibinfo{author}{N.~Shazeer}, \bibinfo{author}{N.~Parmar}, \bibinfo{author}{J.~Uszkoreit}, \bibinfo{author}{L.~Jones}, \bibinfo{author}{A.~N. e.~a. Gomez},
\newblock \bibinfo{title}{Attention is all you need},
\newblock \bibinfo{journal}{Advances in neural information processing systems} \bibinfo{volume}{30} (\bibinfo{year}{2017}) \bibinfo{pages}{5998--6008}.
\bibitem[{Devlin et~al.(2018)Devlin, Chang, Lee, and Toutanova}]{Devlin2018}
\bibinfo{author}{J.~Devlin}, \bibinfo{author}{M.-W. Chang}, \bibinfo{author}{K.~Lee}, \bibinfo{author}{K.~Toutanova},
\newblock \bibinfo{title}{Bert: Pre-training of deep bidirectional transformers for language understanding},
\newblock \bibinfo{journal}{arXiv preprint arXiv:1810.04805}  (\bibinfo{year}{2018}).
\bibitem[{Brown et~al.(2020)Brown, Mann, Ryder, Subbiah, Kaplan, and Dhariwal}]{Brown2020}
\bibinfo{author}{T.~B. Brown}, \bibinfo{author}{B.~Mann}, \bibinfo{author}{N.~Ryder}, \bibinfo{author}{M.~Subbiah}, \bibinfo{author}{J.~Kaplan}, \bibinfo{author}{P.~a.~a. Dhariwal},
\newblock \bibinfo{title}{Language models are few-shot learners},
\newblock \bibinfo{journal}{Advances in neural information processing systems} \bibinfo{volume}{33} (\bibinfo{year}{2020}) \bibinfo{pages}{1877--1901}.
\bibitem[{Radford et~al.(2019)Radford, Wu, Child, Luan, Amodei, and Sutskever}]{Radford2019}
\bibinfo{author}{A.~Radford}, \bibinfo{author}{J.~Wu}, \bibinfo{author}{R.~Child}, \bibinfo{author}{D.~Luan}, \bibinfo{author}{D.~Amodei}, \bibinfo{author}{I.~Sutskever},
\newblock \bibinfo{title}{Language models are unsupervised multitask learners},
\newblock \bibinfo{journal}{OpenAI blog} \bibinfo{volume}{1} (\bibinfo{year}{2019}) \bibinfo{pages}{9}.
\bibitem[{Thirunavukarasu et~al.(2023)Thirunavukarasu, Guntur, and Ghinea}]{Thirunavukarasu2023}
\bibinfo{author}{A.~J. Thirunavukarasu}, \bibinfo{author}{S.~B. Guntur}, \bibinfo{author}{G.~Ghinea},
\newblock \bibinfo{title}{Can chatgpt write a good scientific paper?},
\newblock \bibinfo{journal}{Nature} \bibinfo{volume}{618} (\bibinfo{year}{2023}) \bibinfo{pages}{214--216}.
\bibitem[{Bubeck et~al.(2023)Bubeck, Chandrasekaran, Eldan, Gehrke, Horvitz, and Kamar}]{Bubeck2023}
\bibinfo{author}{S.~Bubeck}, \bibinfo{author}{V.~Chandrasekaran}, \bibinfo{author}{R.~Eldan}, \bibinfo{author}{J.~Gehrke}, \bibinfo{author}{E.~Horvitz}, \bibinfo{author}{E.~e.~a. Kamar},
\newblock \bibinfo{title}{Sparks of artificial general intelligence: Early experiments with gpt-4},
\newblock \bibinfo{journal}{arXiv preprint arXiv:2303.12712}  (\bibinfo{year}{2023}).
\bibitem[{Huang et~al.(2023)Huang, Li, Liu, Wang, and Chen}]{Huang2023}
\bibinfo{author}{C.~Huang}, \bibinfo{author}{S.~Li}, \bibinfo{author}{R.~Liu}, \bibinfo{author}{H.~Wang}, \bibinfo{author}{Y.~Chen},
\newblock \bibinfo{title}{Large foundation models for power systems}  (\bibinfo{year}{2023}). \URLprefix \url{http://arxiv.org/abs/2312.07044}. \DOIprefix\doi{10.1109/PESGM51994.2024.10688670}, \bibinfo{note}{accessed: 2025-02-19}.
\bibitem[{Hamann et~al.(2024)Hamann, Brunschwiler, Gjorgiev, Martins, Puech, and et~al.}]{Hamann2024Foundation}
\bibinfo{author}{H.~F. Hamann}, \bibinfo{author}{T.~Brunschwiler}, \bibinfo{author}{B.~Gjorgiev}, \bibinfo{author}{L.~S.~A. Martins}, \bibinfo{author}{A.~Puech}, \bibinfo{author}{A.~V. et~al.},
\newblock \bibinfo{title}{Foundation models for the electric power grid}  (\bibinfo{year}{2024}). \URLprefix \url{https://arxiv.org/abs/2407.09434v2}. \DOIprefix\doi{10.1016/j.joule.2024.11.002}, \bibinfo{note}{accessed: 2025-02-19}.
\bibitem[{Mohammadabadi et~al.(2024)Mohammadabadi, Entezami, Moghaddam, Orangian, and Nejadshamsi}]{Mohammadabadi2024}
\bibinfo{author}{S.~M.~S. Mohammadabadi}, \bibinfo{author}{M.~Entezami}, \bibinfo{author}{A.~K. Moghaddam}, \bibinfo{author}{M.~Orangian}, \bibinfo{author}{S.~Nejadshamsi},
\newblock \bibinfo{title}{Generative artificial intelligence for distributed learning to enhance smart grid communication},
\newblock \bibinfo{journal}{International Journal of Intelligent Networks} \bibinfo{volume}{5} (\bibinfo{year}{2024}) \bibinfo{pages}{267--274}. \DOIprefix\doi{10.1016/J.IJIN.2024.05.007}.
\bibitem[{Jin and Abhyankar(2024)}]{Jin2024}
\bibinfo{author}{S.~Jin}, \bibinfo{author}{S.~Abhyankar},
\newblock \bibinfo{title}{Chatgrid: Power grid visualization empowered by a large language model},
\newblock \bibinfo{journal}{Proceedings - 2024 IEEE Workshop on Energy Data Visualization, EnergyVis 2024}  (\bibinfo{year}{2024}) \bibinfo{pages}{12--17}. \DOIprefix\doi{10.1109/ENERGYVIS63885.2024.00007}.
\bibitem[{Na et~al.(2024)Na, Yang, Su, Li, Wang, and Chen}]{Na2024}
\bibinfo{author}{Q.~Na}, \bibinfo{author}{Y.~Yang}, \bibinfo{author}{D.~Su}, \bibinfo{author}{X.~Li}, \bibinfo{author}{Y.~Wang}, \bibinfo{author}{Z.~Chen},
\newblock \bibinfo{title}{Speech recognition model inspired on large language model for smart grid dispatching},
\newblock \bibinfo{journal}{ACM International Conference Proceeding Series}  (\bibinfo{year}{2024}) \bibinfo{pages}{439--442}. \URLprefix \url{https://dl.acm.org/doi/10.1145/3674225.3674303}. \DOIprefix\doi{10.1145/3674225.3674303}, \bibinfo{note}{accessed: 2025-02-19}.
\bibitem[{Maddigan and Susnjak(2023)}]{Maddigan2023}
\bibinfo{author}{P.~Maddigan}, \bibinfo{author}{T.~Susnjak},
\newblock \bibinfo{title}{Chat2vis: Generating data visualizations via natural language using chatgpt, codex and gpt-3 large language models},
\newblock \bibinfo{journal}{IEEE Access} \bibinfo{volume}{11} (\bibinfo{year}{2023}) \bibinfo{pages}{45181--45193}. \DOIprefix\doi{10.1109/ACCESS.2023.3274199}.
\bibitem[{Pribylskii and Armeev(2024)}]{Pribylskii2024}
\bibinfo{author}{I.~V. Pribylskii}, \bibinfo{author}{D.~V. Armeev},
\newblock \bibinfo{title}{Ai control system for power grid based on distributed energy resources},
\newblock \bibinfo{journal}{International Conference of Young Specialists on Micro/Nanotechnologies and Electron Devices, EDM}  (\bibinfo{year}{2024}) \bibinfo{pages}{1420--1425}. \DOIprefix\doi{10.1109/EDM61683.2024.10614965}.
\bibitem[{Ruan et~al.(2024)Ruan, Liang, Zhao, Liu, Sun, and et~al.}]{Ruan2024}
\bibinfo{author}{J.~Ruan}, \bibinfo{author}{G.~Liang}, \bibinfo{author}{H.~Zhao}, \bibinfo{author}{G.~Liu}, \bibinfo{author}{X.~Sun}, \bibinfo{author}{J.~Q. et~al.},
\newblock \bibinfo{title}{Applying large language models to power systems: Potential security threats},
\newblock \bibinfo{journal}{IEEE Transactions on Smart Grid} \bibinfo{volume}{15} (\bibinfo{year}{2024}) \bibinfo{pages}{3333--3336}. \DOIprefix\doi{10.1109/TSG.2024.3373256}.
\bibitem[{Li et~al.(2024)Li, Yang, and Sun}]{Li2024}
\bibinfo{author}{J.~Li}, \bibinfo{author}{Y.~Yang}, \bibinfo{author}{J.~Sun},
\newblock \bibinfo{title}{Risks of practicing large language models in smart grid: Threat modeling and validation}  (\bibinfo{year}{2024}). \URLprefix \url{https://arxiv.org/abs/2405.06237v2}, \bibinfo{note}{accessed: 2025-02-19}.
\bibitem[{Selim et~al.(2024)Selim, Zhao, and Yang}]{Selim2024}
\bibinfo{author}{A.~Selim}, \bibinfo{author}{J.~Zhao}, \bibinfo{author}{B.~Yang},
\newblock \bibinfo{title}{Large language model for smart inverter cyber-attack detection via textual analysis of volt/var commands},
\newblock \bibinfo{journal}{IEEE Transactions on Smart Grid}  (\bibinfo{year}{2024}). \DOIprefix\doi{10.1109/TSG.2024.3453648}.
\bibitem[{Zaboli et~al.(2024)Zaboli, Member, Choi, Song, Hong, and Member}]{Zaboli2024Novel}
\bibinfo{author}{A.~Zaboli}, \bibinfo{author}{S.~Member}, \bibinfo{author}{S.~L. Choi}, \bibinfo{author}{T.-J. Song}, \bibinfo{author}{J.~Hong}, \bibinfo{author}{S.~Member},
\newblock \bibinfo{title}{A novel generative ai-based framework for anomaly detection in multicast messages in smart grid communications}  (\bibinfo{year}{2024}). \URLprefix \url{https://arxiv.org/abs/2406.05472v1}, \bibinfo{note}{accessed: 2025-02-19}.
\bibitem[{Zhang et~al.(2024)Zhang, Liu, Sun, Deng, Cheng, and et~al.}]{Zhang2024}
\bibinfo{author}{Z.~Zhang}, \bibinfo{author}{M.~Liu}, \bibinfo{author}{M.~Sun}, \bibinfo{author}{R.~Deng}, \bibinfo{author}{P.~Cheng}, \bibinfo{author}{D.~N. et~al.},
\newblock \bibinfo{title}{Vulnerability of machine learning approaches applied in iot-based smart grid: A review},
\newblock \bibinfo{journal}{IEEE Internet of Things Journal} \bibinfo{volume}{11} (\bibinfo{year}{2024}) \bibinfo{pages}{18951--18975}. \DOIprefix\doi{10.1109/JIOT.2024.3349381}.
\bibitem[{GATTAL(2024)}]{GATTAL2024}
\bibinfo{author}{R.~GATTAL},
\newblock \bibinfo{title}{Llm based approach for anomaly detection in smart grids}  (\bibinfo{year}{2024}). \URLprefix \url{http//localhost:8080/jspui/handle/123456789/11899}, \bibinfo{note}{accessed: 2025-02-19}.
\bibitem[{Lin et~al.(2023)Lin, Zheng, Cai, Fu, Xie, Ma, and Zhang}]{Lin2023}
\bibinfo{author}{C.~Lin}, \bibinfo{author}{Z.~Zheng}, \bibinfo{author}{S.~Cai}, \bibinfo{author}{L.~Fu}, \bibinfo{author}{W.~Xie}, \bibinfo{author}{T.~Ma}, \bibinfo{author}{Z.~Zhang},
\newblock \bibinfo{title}{Knowledge graph completion for power grid main equipment using pretrained language models},
\newblock \bibinfo{journal}{Lecture Notes in Computer Science} \bibinfo{volume}{14089} (\bibinfo{year}{2023}) \bibinfo{pages}{828--838}.
\bibitem[{Chen et~al.(2023)Chen, Wang, Li, Yan, and Cao}]{Chen2023}
\bibinfo{author}{R.~Chen}, \bibinfo{author}{Y.~Wang}, \bibinfo{author}{G.~Li}, \bibinfo{author}{D.~Yan}, \bibinfo{author}{H.~Cao},
\newblock \bibinfo{title}{Pre-training models based knowledge graph multi-hop reasoning for smart grid technology},
\newblock \bibinfo{journal}{Lecture Notes in Electrical Engineering} \bibinfo{volume}{934 LNEE} (\bibinfo{year}{2023}) \bibinfo{pages}{1866--1875}. \URLprefix \url{https://link.springer.com/chapter/10.1007/978-981-19-3998-3-173}. \DOIprefix\doi{10.1007/978-981-19-3998-3-173}, \bibinfo{note}{accessed: 2025-02-19}.
\bibitem[{Yang et~al.(2024)Yang, Li, Zhu, Zheng, Zhang, and Li}]{Yang2024}
\bibinfo{author}{Y.~Yang}, \bibinfo{author}{C.~Li}, \bibinfo{author}{B.~Zhu}, \bibinfo{author}{W.~Zheng}, \bibinfo{author}{F.~Zhang}, \bibinfo{author}{Z.~Li},
\newblock \bibinfo{title}{Enhancing domain-specific text generation for power grid maintenance with p2ft},
\newblock \bibinfo{journal}{Scientific Reports 2024 14:1} \bibinfo{volume}{14} (\bibinfo{year}{2024}) \bibinfo{pages}{1--12}. \URLprefix \url{https://www.nature.com/articles/s41598-024-78078-y}. \DOIprefix\doi{10.1038/s41598-024-78078-y}, \bibinfo{note}{accessed: 2025-02-19}.
\bibitem[{Zhou et~al.(2012)Zhou, Natarajan, Simmhan, and Prasanna}]{Zhou2012}
\bibinfo{author}{Q.~Zhou}, \bibinfo{author}{S.~Natarajan}, \bibinfo{author}{Y.~Simmhan}, \bibinfo{author}{V.~Prasanna},
\newblock \bibinfo{title}{Semantic information modeling for emerging applications in smart grid},
\newblock \bibinfo{journal}{Proceedings of the 9th International Conference on Information Technology, ITNG 2012}  (\bibinfo{year}{2012}) \bibinfo{pages}{775--782}. \DOIprefix\doi{10.1109/ITNG.2012.150}.
\bibitem[{Crapo et~al.(2009)Crapo, Wang, Lizzi, and Larson}]{Crapo2009}
\bibinfo{author}{A.~Crapo}, \bibinfo{author}{X.~Wang}, \bibinfo{author}{J.~Lizzi}, \bibinfo{author}{R.~Larson},
\newblock \bibinfo{title}{The semantically enabled smart grid},
\newblock \bibinfo{journal}{The Road to an Interoperable Grid (Grid-Interop)}  (\bibinfo{year}{2009}).
\bibitem[{Hayawi et~al.(2024)Hayawi, Shahriar, and Al-Ali}]{Hayawi2024}
\bibinfo{author}{K.~Hayawi}, \bibinfo{author}{S.~Shahriar}, \bibinfo{author}{A.~R. Al-Ali},
\newblock \bibinfo{title}{Ethical considerations in ai applications in smart grid}  (\bibinfo{year}{2024}). \URLprefix \url{https://www.preprints.org/manuscript/202412.1883/v1}. \DOIprefix\doi{10.20944/PREPRINTS202412.1883.V1}, \bibinfo{note}{accessed: 2025-02-19}.
\bibitem[{Zhao et~al.(2024)Zhao, Yogarathnam, and Yue}]{Zhao2024}
\bibinfo{author}{T.~Zhao}, \bibinfo{author}{A.~Yogarathnam}, \bibinfo{author}{M.~Yue},
\newblock \bibinfo{title}{A large language model for determining partial tripping of distributed energy resources},
\newblock \bibinfo{journal}{IEEE Transactions on Smart Grid}  (\bibinfo{year}{2024}) \bibinfo{pages}{1--1}. \DOIprefix\doi{10.1109/TSG.2024.3453649}.
\bibitem[{Liu et~al.(2024)Liu, Bai, Wen, Wang, Gu, Liu, Liang, Zhao, and Dong}]{Liu2024}
\bibinfo{author}{G.~Liu}, \bibinfo{author}{Y.~Bai}, \bibinfo{author}{K.~Wen}, \bibinfo{author}{X.~Wang}, \bibinfo{author}{J.~Gu}, \bibinfo{author}{Y.~Liu}, \bibinfo{author}{G.~Liang}, \bibinfo{author}{J.~Zhao}, \bibinfo{author}{Z.~Dong},
\newblock \bibinfo{title}{Lfllm: A robust large language model for short-term load forecasting in smart grids}  (\bibinfo{year}{2024}). \URLprefix \url{https://papers.ssrn.com/abstract=5070252}. \DOIprefix\doi{10.2139/SSRN.5070252}, \bibinfo{note}{accessed: 2025-02-19}.
\bibitem[{Jing and Rahman(2024)}]{Jing2024}
\bibinfo{author}{L.~Jing}, \bibinfo{author}{A.~Rahman},
\newblock \bibinfo{title}{Fault diagnosis in power grids with large language model}  (\bibinfo{year}{2024}). \URLprefix \url{https://arxiv.org/abs/2407.08836v1}, \bibinfo{note}{accessed: 2025-02-19}.
\bibitem[{Wang et~al.(2024)Wang, Chuang, Ke, Chien, Ho, and et~al.}]{Wang2024}
\bibinfo{author}{R.~G. Wang}, \bibinfo{author}{M.~L. Chuang}, \bibinfo{author}{C.~Y. Ke}, \bibinfo{author}{Y.~F. Chien}, \bibinfo{author}{W.~J. Ho}, \bibinfo{author}{K.~C.~C. et~al.},
\newblock \bibinfo{title}{Predicting imminent electrical safety incidents using smart meter big data with large language models},
\newblock \bibinfo{journal}{IEEE Access}  (\bibinfo{year}{2024}). \DOIprefix\doi{10.1109/ACCESS.2024.3514209}.
\bibitem[{Nguyen et~al.(2024)Nguyen, Nguyen, Hwang, Kuo, Chen, and et~al.}]{Nguyen2024}
\bibinfo{author}{L.-H. Nguyen}, \bibinfo{author}{V.-L. Nguyen}, \bibinfo{author}{R.-H. Hwang}, \bibinfo{author}{J.-J. Kuo}, \bibinfo{author}{Y.-W. Chen}, \bibinfo{author}{C.-C.~H. et~al.},
\newblock \bibinfo{title}{Towards secured smart grid 2.0: Exploring security threats, protection models, and challenges}  (\bibinfo{year}{2024}). \URLprefix \url{http://arxiv.org/abs/2411.04365}. \DOIprefix\doi{10.1109/COMST.2024.3493630}, \bibinfo{note}{accessed: 2025-02-19}.
\bibitem[{Shi et~al.(2024)Shi, Fang, Chen, Gu, Ma, and et~al.}]{Shi2024}
\bibinfo{author}{H.~Shi}, \bibinfo{author}{L.~Fang}, \bibinfo{author}{X.~Chen}, \bibinfo{author}{C.~Gu}, \bibinfo{author}{K.~Ma}, \bibinfo{author}{X.~Z. et~al.},
\newblock \bibinfo{title}{Review of the opportunities and challenges to accelerate mass-scale application of smart grids with large-language models},
\newblock \bibinfo{journal}{IET Smart Grid} \bibinfo{volume}{7} (\bibinfo{year}{2024}) \bibinfo{pages}{737--759}. \DOIprefix\doi{10.1049/STG2.12191}, \bibinfo{note}{accessed: 2025-02-19}.
\bibitem[{Bui et~al.(2024)Bui, Das, Hussain, Hollweg, and Su}]{Bui2024}
\bibinfo{author}{V.-H. Bui}, \bibinfo{author}{S.~Das}, \bibinfo{author}{A.~Hussain}, \bibinfo{author}{G.~V. Hollweg}, \bibinfo{author}{W.~Su},
\newblock \bibinfo{title}{A critical review of safe reinforcement learning techniques in smart grid applications}  (\bibinfo{year}{2024}). \URLprefix \url{https://arxiv.org/abs/2409.16256v1}, \bibinfo{note}{accessed: 2025-02-19}.
\bibitem[{Wilker et~al.(2017)Wilker, Meisel, and Sauter}]{Wilker2017}
\bibinfo{author}{S.~Wilker}, \bibinfo{author}{M.~Meisel}, \bibinfo{author}{T.~Sauter},
\newblock \bibinfo{title}{Smart grid architecture model standardization and the applicability of domain language specific modeling tools},
\newblock \bibinfo{journal}{IEEE International Symposium on Industrial Electronics}  (\bibinfo{year}{2017}) \bibinfo{pages}{152--157}. \DOIprefix\doi{10.1109/ISIE.2017.8001239}, \bibinfo{note}{accessed: 2025-02-19}.
\bibitem[{Zibaeirad et~al.(2024)Zibaeirad, Koleini, Bi, Hou, and Wang}]{Zibaeirad2024}
\bibinfo{author}{A.~Zibaeirad}, \bibinfo{author}{F.~Koleini}, \bibinfo{author}{S.~Bi}, \bibinfo{author}{T.~Hou}, \bibinfo{author}{T.~Wang},
\newblock \bibinfo{title}{A comprehensive survey on the security of smart grid: Challenges, mitigations, and future research opportunities}  (\bibinfo{year}{2024}). \URLprefix \url{https://arxiv.org/abs/2407.07966v1}, \bibinfo{note}{accessed: 2025-02-19}.
\bibitem[{Escobar et~al.(2021)Escobar, Matamoros, Padilla, Reyes, and Espinosa}]{Escobar2021}
\bibinfo{author}{J.~J.~M. Escobar}, \bibinfo{author}{O.~M. Matamoros}, \bibinfo{author}{R.~T. Padilla}, \bibinfo{author}{I.~L. Reyes}, \bibinfo{author}{H.~Q. Espinosa},
\newblock \bibinfo{title}{A comprehensive review on smart grids: Challenges and opportunities},
\newblock \bibinfo{journal}{Sensors} \bibinfo{volume}{21} (\bibinfo{year}{2021}) \bibinfo{pages}{6978}. \URLprefix \url{https://doi.org/10.3390/s21216978}. \DOIprefix\doi{10.3390/s21216978}, \bibinfo{note}{accessed: 2025-02-19}.
\bibitem[{Li et~al.(2022)Li, Wei, Li, Dong, and Shahidehpour}]{Li2022}
\bibinfo{author}{Y.~Li}, \bibinfo{author}{X.~Wei}, \bibinfo{author}{Y.~Li}, \bibinfo{author}{Z.~Dong}, \bibinfo{author}{M.~Shahidehpour},
\newblock \bibinfo{title}{Detection of false data injection attacks in smart grid: A secure federated deep learning approach},
\newblock \bibinfo{journal}{{IEEE} Transactions on Smart Grid} \bibinfo{volume}{13} (\bibinfo{year}{2022}) \bibinfo{pages}{4862--4872}. \URLprefix \url{https://doi.org/10.1109/tsg.2022.3204796}. \DOIprefix\doi{10.1109/TSG.2022.3204796}, \bibinfo{note}{accessed: 2025-02-19}.
\bibitem[{Rottondi et~al.(2013)Rottondi, Verticale, and Krau{\ss}}]{Rottondi2013}
\bibinfo{author}{C.~Rottondi}, \bibinfo{author}{G.~Verticale}, \bibinfo{author}{C.~Krau{\ss}},
\newblock \bibinfo{title}{Distributed privacy-preserving aggregation of metering data in smart grids},
\newblock \bibinfo{journal}{{IEEE} Journal on Selected Areas in Communications} \bibinfo{volume}{31} (\bibinfo{year}{2013}) \bibinfo{pages}{1342--1354}. \URLprefix \url{https://doi.org/10.1109/jsac.2013.130716}. \DOIprefix\doi{10.1109/JSAC.2013.130716}, \bibinfo{note}{accessed: 2025-02-19}.
\bibitem[{Munshi and Mohamed(2018)}]{Munshi2018}
\bibinfo{author}{A.~A. Munshi}, \bibinfo{author}{Y.~Mohamed},
\newblock \bibinfo{title}{Data lake lambda architecture for smart grids big data analytics},
\newblock \bibinfo{journal}{{IEEE} Access} \bibinfo{volume}{6} (\bibinfo{year}{2018}) \bibinfo{pages}{40463--40471}. \URLprefix \url{https://doi.org/10.1109/access.2018.2858256}. \DOIprefix\doi{10.1109/ACCESS.2018.2858256}, \bibinfo{note}{accessed: 2025-02-19}.
\bibitem[{Alladi et~al.(2019)Alladi, Chamola, Rodrigues, and Kozlov}]{Alladi2019}
\bibinfo{author}{T.~Alladi}, \bibinfo{author}{V.~Chamola}, \bibinfo{author}{J.~J. P.~C. Rodrigues}, \bibinfo{author}{S.~A. Kozlov},
\newblock \bibinfo{title}{Blockchain in smart grids: A review on different use cases},
\newblock \bibinfo{journal}{Sensors} \bibinfo{volume}{19} (\bibinfo{year}{2019}) \bibinfo{pages}{4862}. \URLprefix \url{https://doi.org/10.3390/s19224862}. \DOIprefix\doi{10.3390/s19224862}, \bibinfo{note}{accessed: 2025-02-19}.
\bibitem[{Li et~al.(2024)Li, Yang, and Sun}]{li_risks_2024}
\bibinfo{author}{J.~Li}, \bibinfo{author}{Y.~Yang}, \bibinfo{author}{J.~Sun}, \bibinfo{title}{Risks of {Practicing} {Large} {Language} {Models} in {Smart} {Grid}: {Threat} {Modeling} and {Validation}}, \bibinfo{year}{2024}. \URLprefix \url{http://arxiv.org/abs/2405.06237}. \DOIprefix\doi{10.48550/arXiv.2405.06237}, \bibinfo{note}{accessed: 2025-02-19}.
\bibitem[{Ayyamperumal and Ge(2024)}]{ayyamperumal_current_2024}
\bibinfo{author}{S.~G. Ayyamperumal}, \bibinfo{author}{L.~Ge}, \bibinfo{title}{Current state of {LLM} {Risks} and {AI} {Guardrails}}, \bibinfo{year}{2024}. \URLprefix \url{http://arxiv.org/abs/2406.12934}. \DOIprefix\doi{10.48550/arXiv.2406.12934}, \bibinfo{note}{accessed: 2025-02-19}.
\bibitem[{Bhattarai et~al.(2019)Bhattarai, Paudyal, Luo, Mohanpurkar, Cheung, and et~al.}]{Bhattarai2019}
\bibinfo{author}{B.~Bhattarai}, \bibinfo{author}{S.~Paudyal}, \bibinfo{author}{Y.~Luo}, \bibinfo{author}{M.~Mohanpurkar}, \bibinfo{author}{K.~Cheung}, \bibinfo{author}{R.~T. et~al.},
\newblock \bibinfo{title}{Big data analytics in smart grids: state‐of‐the‐art, challenges, opportunities, and future directions},
\newblock \bibinfo{journal}{IET Smart Grid} \bibinfo{volume}{13} (\bibinfo{year}{2019}) \bibinfo{pages}{1--22}. \URLprefix \url{https://doi.org/10.1049/iet-stg.2018.0261}. \DOIprefix\doi{10.1049/iet-stg.2018.0261}, \bibinfo{note}{accessed: 2025-02-19}.
\bibitem[{Cao et~al.(2017)Cao, Lin, Song, Zhang, and Wang}]{Cao2017}
\bibinfo{author}{Z.~Cao}, \bibinfo{author}{J.~Lin}, \bibinfo{author}{Y.~Song}, \bibinfo{author}{Y.~Zhang}, \bibinfo{author}{X.~Wang},
\newblock \bibinfo{title}{Optimal cloud computing resource allocation for demand side management in smart grid},
\newblock \bibinfo{journal}{{IEEE} Transactions on Smart Grid} \bibinfo{volume}{8} (\bibinfo{year}{2017}) \bibinfo{pages}{1943--1955}. \URLprefix \url{https://doi.org/10.1109/tsg.2015.2512712}. \DOIprefix\doi{10.1109/TSG.2015.2512712}, \bibinfo{note}{accessed: 2025-02-19}.
\bibitem[{Zeid et~al.(2019)Zeid, Sundaram, Moghaddam, Kamarthi, and Marion}]{Zeid2019}
\bibinfo{author}{A.~Zeid}, \bibinfo{author}{S.~Sundaram}, \bibinfo{author}{M.~Moghaddam}, \bibinfo{author}{S.~Kamarthi}, \bibinfo{author}{T.~Marion},
\newblock \bibinfo{title}{Interoperability in smart manufacturing: Research challenges},
\newblock \bibinfo{journal}{Machines} \bibinfo{volume}{7} (\bibinfo{year}{2019}) \bibinfo{pages}{21}. \URLprefix \url{https://doi.org/10.3390/machines7020021}. \DOIprefix\doi{10.3390/machines7020021}, \bibinfo{note}{accessed: 2025-02-19}.
\bibitem[{Cintuglu et~al.(2018)Cintuglu, Youssef, and Mohammed}]{Cintuglu2018}
\bibinfo{author}{M.~H. Cintuglu}, \bibinfo{author}{T.~Youssef}, \bibinfo{author}{O.~A. Mohammed},
\newblock \bibinfo{title}{Development and application of a real-time testbed for multiagent system interoperability: A case study on hierarchical microgrid control},
\newblock \bibinfo{journal}{{IEEE} Transactions on Smart Grid} \bibinfo{volume}{9} (\bibinfo{year}{2018}) \bibinfo{pages}{1759--1768}. \URLprefix \url{https://doi.org/10.1109/tsg.2016.2599265}. \DOIprefix\doi{10.1109/TSG.2016.2599265}, \bibinfo{note}{accessed: 2025-02-19}.
\bibitem[{Sun et~al.(2019)Sun, Guo, Qi, Ajjarapu, Bravo, and et~al.}]{Sun2019}
\bibinfo{author}{H.~Sun}, \bibinfo{author}{Q.~Guo}, \bibinfo{author}{J.~Qi}, \bibinfo{author}{V.~Ajjarapu}, \bibinfo{author}{R.~Bravo}, \bibinfo{author}{J.~C. et~al.},
\newblock \bibinfo{title}{Review of challenges and research opportunities for voltage control in smart grids},
\newblock \bibinfo{journal}{{IEEE} Transactions on Power Systems} \bibinfo{volume}{34} (\bibinfo{year}{2019}) \bibinfo{pages}{2790--2801}. \URLprefix \url{https://doi.org/10.1109/tpwrs.2019.2897948}. \DOIprefix\doi{10.1109/TPWRS.2019.2897948}, \bibinfo{note}{accessed: 2025-02-19}.
\bibitem[{Farraj et~al.(2018)Farraj, Hammad, and Kundur}]{Farraj2018}
\bibinfo{author}{A.~K. Farraj}, \bibinfo{author}{E.~M. Hammad}, \bibinfo{author}{D.~Kundur},
\newblock \bibinfo{title}{A cyber-physical control framework for transient stability in smart grids},
\newblock \bibinfo{journal}{{IEEE} Transactions on Smart Grid} \bibinfo{volume}{9} (\bibinfo{year}{2018}) \bibinfo{pages}{1205--1215}. \URLprefix \url{https://doi.org/10.1109/tsg.2016.2581588}. \DOIprefix\doi{10.1109/TSG.2016.2581588}, \bibinfo{note}{accessed: 2025-02-19}.
\bibitem[{Ayad et~al.(2018)Ayad, Farag, Youssef, and El-Saadany}]{Ayad2018}
\bibinfo{author}{A.~Ayad}, \bibinfo{author}{H.~Farag}, \bibinfo{author}{A.~Youssef}, \bibinfo{author}{E.~El-Saadany},
\newblock \bibinfo{title}{Detection of false data injection attacks in smart grids using recurrent neural networks},
\newblock in: \bibinfo{booktitle}{2018 {IEEE} Power \& Energy Society Innovative Smart Grid Technologies Conference ({ISGT})}, \bibinfo{year}{2018}, pp. \bibinfo{pages}{1--5}. \URLprefix \url{https://doi.org/10.1109/isgt.2018.8403355}. \DOIprefix\doi{10.1109/ISGT.2018.8403355}, \bibinfo{note}{accessed: 2025-02-19}.
\bibitem[{Farzaneh et~al.(2021)Farzaneh, Malehmirchegini, Bejan, Afolabi, Mulumba, and Daka}]{Farzaneh2021}
\bibinfo{author}{H.~Farzaneh}, \bibinfo{author}{L.~Malehmirchegini}, \bibinfo{author}{A.~Bejan}, \bibinfo{author}{T.~Afolabi}, \bibinfo{author}{A.~Mulumba}, \bibinfo{author}{P.~P. Daka},
\newblock \bibinfo{title}{Artificial intelligence evolution in smart buildings for energy efficiency},
\newblock \bibinfo{journal}{Applied Sciences} \bibinfo{volume}{11} (\bibinfo{year}{2021}) \bibinfo{pages}{763}.
\bibitem[{Solyali(2020)}]{Solyali2020}
\bibinfo{author}{D.~Solyali},
\newblock \bibinfo{title}{A comparative analysis of machine learning approaches for short-/long-term electricity load forecasting in cyprus},
\newblock \bibinfo{journal}{Sustainability} \bibinfo{volume}{12} (\bibinfo{year}{2020}) \bibinfo{pages}{3612}.
\bibitem[{Lin et~al.(2022)Lin, Xia, Cao, Meng, Zhang, Feng, Zhao, and Wang}]{Lin2022}
\bibinfo{author}{S.~Lin}, \bibinfo{author}{L.~Xia}, \bibinfo{author}{Y.~Cao}, \bibinfo{author}{H.-L. Meng}, \bibinfo{author}{L.~Zhang}, \bibinfo{author}{J.-J. Feng}, \bibinfo{author}{Y.~Zhao}, \bibinfo{author}{A.-J. Wang},
\newblock \bibinfo{title}{Electronic regulation of znco dual-atomic active sites entrapped in 1d@2d hierarchical n-doped carbon for efficient synergistic catalysis of oxygen reduction in zn-air battery},
\newblock \bibinfo{journal}{Small}  (\bibinfo{year}{2022}) \bibinfo{pages}{e2107141}.
\bibitem[{Nourani et~al.(2019)Nourani, Elkiran, Abdullahi, and Tahsin}]{Nourani2019}
\bibinfo{author}{V.~Nourani}, \bibinfo{author}{G.~Elkiran}, \bibinfo{author}{J.~Abdullahi}, \bibinfo{author}{A.~Tahsin},
\newblock \bibinfo{title}{Multi-region modeling of daily global solar radiation with artificial intelligence ensemble},
\newblock \bibinfo{journal}{Natural Resources Research}  (\bibinfo{year}{2019}) \bibinfo{pages}{1--22}.
\bibitem[{Wang et~al.(2020)Wang, Ogbodo, Huang, Qiu, Hisada, and Abdallah}]{Wang2020}
\bibinfo{author}{Z.~Wang}, \bibinfo{author}{M.~Ogbodo}, \bibinfo{author}{H.~Huang}, \bibinfo{author}{C.~Qiu}, \bibinfo{author}{M.~Hisada}, \bibinfo{author}{A.~B. Abdallah},
\newblock \bibinfo{title}{Aebis: Ai-enabled blockchain-based electric vehicle integration system for power management in smart grid platform},
\newblock \bibinfo{journal}{{IEEE} Access} \bibinfo{volume}{8} (\bibinfo{year}{2020}) \bibinfo{pages}{226409--226421}.
\bibitem[{Kuzlu et~al.(2020)Kuzlu, Cali, Sharma, and G{\"u}ler}]{Kuzlu2020}
\bibinfo{author}{M.~Kuzlu}, \bibinfo{author}{U.~Cali}, \bibinfo{author}{V.~Sharma}, \bibinfo{author}{{\"Ozg{\"u}r}.~G{\"u}ler},
\newblock \bibinfo{title}{Gaining insight into solar photovoltaic power generation forecasting utilizing explainable artificial intelligence tools},
\newblock \bibinfo{journal}{{IEEE} Access} \bibinfo{volume}{8} (\bibinfo{year}{2020}) \bibinfo{pages}{187814--187823}.
\bibitem[{Kushawaha(2024)}]{Kushawaha2024}
\bibinfo{author}{V.~Kushawaha},
\newblock \bibinfo{title}{Enhancing energy efficiency: advances in smart grid optimization},
\newblock \bibinfo{journal}{International Journal of Innovative Research in Engineering \& Management}  (\bibinfo{year}{2024}) \bibinfo{pages}{100--105}.
\bibitem[{Olatunde(2024)}]{Olatunde2024}
\bibinfo{author}{T.~Olatunde},
\newblock \bibinfo{title}{The impact of smart grids on energy efficiency: a comprehensive review},
\newblock \bibinfo{journal}{Engineering Science \& Technology Journal}  (\bibinfo{year}{2024}) \bibinfo{pages}{1257--1269}.
\bibitem[{Maurya(2024)}]{Maurya2024}
\bibinfo{author}{P.~Maurya},
\newblock \bibinfo{title}{Artificial intelligence to enhance energy management and distribution in smart grid communication networks},
\newblock \bibinfo{journal}{tjjpt}  (\bibinfo{year}{2024}) \bibinfo{pages}{4366--4378}.
\bibitem[{Marques(2024)}]{Marques2024}
\bibinfo{author}{P.~Marques}, \bibinfo{title}{Artificial intelligence technologies applied to smart grids and management}, \bibinfo{howpublished}{\url{https://doi.org/10.20944/preprints202406.1248.v1}}, \bibinfo{year}{2024}. \bibinfo{note}{Accessed: 2025-02-19}.
\bibitem[{Mohammed(2024)}]{Mohammed2024}
\bibinfo{author}{S.~Mohammed},
\newblock \bibinfo{title}{A review on the evaluation of feature selection using machine learning for cyber-attack detection in smart grid},
\newblock \bibinfo{journal}{Ieee Access} \bibinfo{volume}{12} (\bibinfo{year}{2024}) \bibinfo{pages}{44023--44042}.
\bibitem[{Bouramdane(2023)}]{Bouramdane2023}
\bibinfo{author}{A.~Bouramdane},
\newblock \bibinfo{title}{Cyberattacks in smart grids: challenges and solving the multi-criteria decision-making for cybersecurity options, including ones that incorporate artificial intelligence, using an analytical hierarchy process},
\newblock \bibinfo{journal}{Journal of Cybersecurity and Privacy} \bibinfo{volume}{3} (\bibinfo{year}{2023}) \bibinfo{pages}{662--705}.
\bibitem[{Manas(2015)}]{Manas2015}
\bibinfo{author}{M.~Manas},
\newblock \bibinfo{title}{Analysis of design of technologies, tariff structures, and regulatory policies for sustainable growth of the smart grid},
\newblock \bibinfo{journal}{Energy Technology \& Policy} \bibinfo{volume}{2} (\bibinfo{year}{2015}) \bibinfo{pages}{28--38}.
\bibitem[{Chung and Zhang(2023)}]{Chung2023}
\bibinfo{author}{S.~Chung}, \bibinfo{author}{Y.~Zhang},
\newblock \bibinfo{title}{Artificial intelligence applications in electric distribution systems: post-pandemic progress and prospect},
\newblock \bibinfo{journal}{Applied Sciences} \bibinfo{volume}{13} (\bibinfo{year}{2023}) \bibinfo{pages}{6937}.
\bibitem[{Ezeigweneme(2024)}]{Ezeigweneme2024}
\bibinfo{author}{C.~Ezeigweneme},
\newblock \bibinfo{title}{Smart grids in industrial paradigms: a review of progress, benefits, and maintenance implications: analyzing the role of smart grids in predictive maintenance and the integration of renewable energy sources, along with their overall impact on the industri},
\newblock \bibinfo{journal}{Engineering Science \& Technology Journal} \bibinfo{volume}{5} (\bibinfo{year}{2024}) \bibinfo{pages}{1--20}.

\end{thebibliography}





\appendix

\section{Illustrative Examples for LLM Use Cases in Smart Grids}\label{appendix}

This appendix aims to bring to life the 30 distinct use cases of LLMs in smart grids, which are elaborated in the main paper. Instead of a list, we will walk through these applications with examples that highlight their practical value and potential impact across various areas of the smart grid.

\bigskip
\textbf{Cluster 1: Intelligent Grid Operations \& Management}
\bigskip

Imagine a grid operator facing a severe heatwave. Instead of manually adjusting settings, they can simply ask the LLM agent: "Increase cooling capacity to residential areas by prioritizing power to those substations." The LLM agent interprets this request and automatically reconfigures the grid, boosting power flow to the necessary substations and ensuring adequate cooling during peak demand.

Consider a scenario where the LLM agent spots unusual network activity targeting a substation control system. The LLM agent immediately flags: "We have detected a surge of traffic coming from an unknown IP address, X.X.X.X, directed at the control server in Substation Beta. This pattern looks like a potential cyberattack." This immediate alert provides a clear picture of the threat, allowing for a rapid response.

In another instance, when a severe weather warning arrives, the LLM agent proactively generates a resilience scenario. It says: "Expect extreme snowfall and wind. We anticipate downed power lines and substation outages in rural areas initially. This could lead to isolated communities experiencing power disruptions for 48-72 hours, affecting up to 10,000 customers, due to limited repair access." This narrative prediction aids in planning a more effective response.

When a customer reports an outage and sensors trigger alarms, the LLM agent analyzes the situation and offers a solution. It states: "The fault appears to be on Distribution Feeder F7 in Zone 3 based on outage reports and ‘last gasp’ signals. The recommended action is: First, dispatch a line crew to the area. Next, isolate Feeder F7 for safety. Conduct a thorough inspection and repair. We estimate restoration in 3 hours." This provides a clear workflow for prompt repairs.

Finally, imagine a microgrid operator who wants to optimize renewable energy usage. They can instruct the LLM agent: "Prioritize renewable energy sources for the next six hours while keeping the battery charge above 50\%." The LLM agent then optimizes the microgrid's operations, balancing solar, wind, and battery storage to achieve maximum renewable energy use within constraints.

\bigskip
\textbf{Cluster 2: Smart Energy Markets \& Trading}
\bigskip

LLM agents are transforming the energy market as well. For instance, a renewable energy producer can have an LLM agent autonomously negotiate a PPA with a utility. The LLM agent considers market prices, demand forecasts, and regulations to propose terms like a fixed price per kWh, a 15-year contract, and the transfer of renewable energy certificates.

Furthermore, LLM agents can perform sentiment analysis. The LLM agent examines news and social media, identifies a negative trend towards natural gas prices due to global instability and then predicts a short-term price increase, enabling a more informed trading strategy.

Decentralized energy trading is another promising area. A homeowner selling excess solar energy can post an offer, and a local business looking to buy renewable energy can post their request. The LLM agent matches these offers, fostering direct peer-to-peer energy transactions within the community.

Finally, consider an LLM agent dynamically adjusting prices during an unexpected heatwave. When the LLM agent detects a surge in demand for air conditioning, it implements a temporary peak-hour surcharge to reduce demand and prevent grid overload, ensuring stable service.

\bigskip
\textbf{Cluster 3: Personalized Energy Management \& Customer Engagement}
\bigskip

LLM agents provide personalized energy saving recommendations to customers. An LLM agent chatbot analyzes a home's energy use and suggests: "Based on your consumption, reducing your pool pump's runtime by two hours could save you around \$20 a month. Would you like to see smart timer options for your pool pump?"

In case of a power outage, the LLM agent proactively communicates with affected customers. It might send notifications stating: "There is a power outage in your area due to equipment issues. Crews are working to restore power, estimated by 8 PM. For outage safety tips, visit [link]."

LLM agents can also offer personalized tariff recommendations based on consumption patterns. For example, an LLM agent can say: "We highly recommend the 'EV Saver' plan for you, given your EV charging patterns. It offers lower rates overnight, potentially saving you \$35 per month on your bill."

Integration with smart homes is another application. A user can command via voice: “Set the thermostat to 72 degrees and dim the living room lights to 50\%.” The LLM agent seamlessly processes this command and adjusts the smart home devices accordingly.

\bigskip
\textbf{Cluster 4: Smart Grid Planning \& Education}
\bigskip

LLM agents are used for the creation of digital twins of grid infrastructure. By analyzing CAD drawings and real-time sensor data, an LLM agent can create a detailed, 3D model of a new substation for operational monitoring and maintenance.

Engineers can also use LLM agents to generate innovative network designs. Given constraints such as: "design a distribution network for a new community prioritizing underground cabling, high renewable energy penetration, and minimal visual impact”, the LLM agent can generate various network options optimizing layouts, cable routing, and substation configurations.

Public education is enhanced through LLM agents. For homeowners, the LLM agent can create an engaging video explaining smart meters, highlighting benefits such as energy tracking and time-of-use billing, in simple and clear terms.

Interactive training for grid operators is another key application, enabling trainees to simulate and ask, "Show me the load flow on Feeder Alpha under peak demand conditions." The LLM agent provides dynamic visualizations of load flow, giving trainees the ability to understand the real grid behavior.

\bigskip
\textbf{Cluster 5: Smart Grid Security \& Compliance}
\bigskip

In terms of security, LLM agents can detect cyberthreats proactively. Analyzing firewall logs, an LLM agent might generate a high-priority alert, stating: "We’ve detected brute-force login attempts on a critical database server, DB-GridControl, originating from multiple countries. This indicates a distributed attack. We recommend an immediate investigation."

LLM agents can also automate regulatory compliance. When a new NERC CIP standard is released, the LLM agent can analyze it and generate a report outlining required asset and security procedure updates, facilitating proactive compliance.

Furthermore, LLM agents can perform vulnerability assessments by analyzing PLC firmware. If the LLM agent identifies a potential buffer overflow, it can recommend upgrading to the latest firmware and implementing input validation as a preventative measure.

\bigskip
\textbf{Cluster 6: Advanced Data Analysis \& Knowledge Discovery}
\bigskip

LLM agents can generate narratives to help understand grid events. Analyzing data during a storm, an LLM agent can say: "The outage cascade started with high winds causing tree contact with distribution lines in Zone 7, which led to outages. Voltage fluctuations then triggered protective relays at Substation Delta, resulting in a larger outage affecting Zones 7, 8, and 9.”

The LLM agent also facilitates cross-domain knowledge integration. Analyzing cybersecurity and telecom research papers, it might propose adapting a novel intrusion detection method used in mobile networks to enhance smart grid network security.

LLM agents go further than reporting raw data; they offer contextual insight. Instead of saying “Voltage at Busbar 3: 245 kV,” it might state: “The voltage at Busbar 3 (245 kV) is slightly elevated compared to 230 kV, which is within limits, yet unusual given the low load conditions, warranting monitoring of the voltage regulator."

\bigskip
\textbf{Cluster 7: Emerging Applications \& Societal Impact}
\bigskip

LLM agents are being applied for environmental justice analysis. By analyzing outage, demographic, and industrial data, an LLM agent may find: "Low-income communities in District 4 experience more frequent and longer outages (30\% and 40\%, respectively) compared to wealthier districts. These communities also are near older, polluting power plants, highlighting environmental justice concerns relating to grid reliability.”

In assessing social impact of grid projects, the LLM agent can analyze public feedback. When reviewing public opinion of a wind farm project, an LLM agent might find: "Sentiment analysis shows significant community concern about visual impact and noise. Addressing these concerns through community engagement and exploring mitigation options is key for the project's acceptance."

LLM agents can generate a comprehensive ‘biography’ of grid assets, which can support life-cycle management. For example, it can summarize the life of a transformer: "Transformer TX-52: Installed in 1995; major maintenance in 2010 (oil replacement) and 2018 (bushing replacement); current sensor data indicates a gradual temperature increase and minor insulation degradation. Inspection reports from 2023 indicate minor corrosion. Predicted remaining useful life: 7 years; enhanced monitoring is recommended."

LLM agents also enhance cross-lingual communications during grid emergencies. During a wildfire, the LLM agent can translate safety instructions and alerts into multiple languages, ensuring broader reach.

\bigskip
\textbf{Cluster 8: LLM-Enhanced RL in Smart Grids}
\bigskip

A grid operator, concerned about the system's response to volatile renewable energy input, instructs the LLM agent: "Adjust the RL agent to prioritize grid frequency stability over cost for the next hour.” The LLM agent analyzes the code implementation of the RL algorithm, specifically modifying its reward function to penalize frequency deviations more heavily. It also constrains the action space of the RL agent, limiting the extent to which it can reduce renewable energy output if grid frequency becomes unstable. The LLM agent confirms the change with a message: "The RL algorithm has been adjusted to prioritize frequency stability, with an increased penalty for deviations and constrained output control for renewable resources. Monitoring will continue.”

LLM agents can also generate training scenarios in high level and then translate these scenarios into numerical vectors to feed into RL agents. It could generate: “Scenario: Geomagnetic disturbance causing widespread voltage instability,” to ensure the RL agent can handle diverse challenges and has the necessary resilience.

Finally, LLM agents can model prosumer behavior to enhance RL training. During demand response events, the LLM agent predicts prosumer responses, considering price sensitivity, time of day, and weather conditions, creating a more realistic training environment. LLM agents could be generated based on the variation in the customer profiles to represent their behavior and aspirations.

These examples show how LLM agents can be a powerful force in creating more intelligent, efficient, and responsive smart grids.

\end{document}